\documentclass[12pt]{article} 
\usepackage{graphics,graphicx}
\usepackage[lofdepth,lotdepth]{subfig}
\usepackage[polutonikogreek,english]{babel}
\usepackage{float}
\usepackage{color}
\usepackage{latexsym}
\usepackage{amsmath}
\usepackage{amssymb}
\usepackage{pifont}
\usepackage{rotating}
\usepackage{pdflscape}
\usepackage{microtype}

\begin{document} 
\title{Hadron energy response of the Iron Calorimeter detector at the
India-based Neutrino Observatory}  
 
\author{
{\small Moon~Moon~Devi$^{1}\footnote{moonmoon4u@tifr.res.in}$} , 
{\small Anushree~Ghosh$^{2}\footnote{anushree@hri.res.in}$} ,
{\small Daljeet~Kaur$^{3}\footnote{daljeet.kaur97@gmail.com}$} ,
 {\small Lakshmi~S.~Mohan$^{4}\footnote{slakshmi@imsc.res.in}$} ,
\and 
{\small Sandhya~Choubey$^{2}$},
{\small Amol~Dighe$^{1}$}, 
{\small D.~Indumathi$^{4}$}, 
{\small Sanjeev~Kumar$^{3}$},
\and 
{\small M.~V.~N.~Murthy$^{4}$},  
{\small Md.~Naimuddin$^{3}$}  \\
\\ 
\vspace{0.05cm} 
\and 
 $^{1}${\it \small Tata Institute of Fundamental Research, 
Mumbai 400 005, India} \\ 
 \and  
$^{2}${\it \small Harishchandra Research Institute, 
Allahabad 211 002, India}\\ 
 \and  
$^{3}${\it \small Department of Physics and Astrophysics, 
Delhi University, New Delhi 110 007, India}\\ 
 \and  
$^{4}${\it \small The Institute of Mathematical Sciences, 
Chennai 600 113, India} 
 } 
\maketitle  
 
\begin{abstract} The results of a Monte Carlo simulation study 
of the hadron energy response for the magnetized Iron CALorimeter 
detector, ICAL, proposed to be located at the India-based Neutrino 
Observatory (INO) is presented. Using a GEANT4 modeling of the detector 
ICAL, interactions of atmospheric neutrinos with target nuclei are simulated. 
The detector response to hadrons propagating through it is
investigated using the hadron hit multiplicity in the active detector elements.
The detector response to charged pions of fixed energy is studied first, 
followed by the average response to the hadrons produced in atmospheric 
neutrino interactions using events simulated with the NUANCE event generator. 
The shape of the hit distribution is observed to fit the
Vavilov distribution, which reduces to a Gaussian at high energies. 
In terms of the parameters of this distribution, we present 
the hadron energy resolution as a function of hadron energy, and 
the calibration of hadron energy as a function of the hit multiplicity.
The energy resolution for hadrons is found to be in the range
85\% (for 1GeV) -- 36\% (for 15 GeV).

\end{abstract} 
 
\newpage

\section{Introduction} 
 \label{intro}
The India-based Neutrino Observatory (INO) \cite{Athar:2006yb} is a 
planned underground laboratory in Southern part of India. 
The primary focus in the first phase of the program is to study 
atmospheric neutrinos with a magnetized 
iron calorimeter (ICAL) detector similar in concept to the design of 
the MONOLITH detector \cite{TabarellideFatis:2001wy}.  The detector is 
designed to observe neutrino (and anti-neutrino) interactions in the 
GeV range.  Recent measurements of the mixing angle $\theta_{13}$ in 
reactor experiments \cite{An:2012eh,Ahn:2012nd,Abe:2011fz} 
will enable ICAL to pin down the mass ordering of neutrinos by separate 
measurements of $\nu_\mu$ and $\bar{\nu}_\mu$ interactions, exploiting 
the matter effects in the Earth. 
 
The magnetized ICAL detector will consist of three identical modules 
with 151 layers of iron plates interspersed with Resistive 
Plate Chamber (RPC) detectors, and will be approximately 50 kt in mass.  
Atmospheric neutrinos (anti-neutrinos) interact with the iron target 
in the detector through quasi-elastic (QE), resonance (RS) and deep 
inelastic scattering (DIS) processes, producing charged leptons in 
charged-current (CC) interactions, with a set of possible final hadrons, 
typically one pion and a nucleon in RS, and multiple hadrons in 
DIS interactions.  At these energies, coherent interactions and the 
interactions with the electrons in the detector are rare and are ignored. 
ICAL is expected to have good charge identification efficiency and good 
tracking and energy resolution for muons produced in CC interactions.
The reach of ICAL for determining the mass hierarchy of neutrinos and 
measuring the atmospheric mixing parameters using only the information 
on muon energy and direction has been presented in \cite{anushree-1,tarak-1}.

While the information on muon energy and direction is crucial for
the physics goals of ICAL, the detector is also sensitive to hadrons,
and the additional information from the response of the detector to 
hadrons can only enhance its physics reach. 
Recently it has been shown that with the inclusion of hadron energy 
information, the analysis of atmospheric neutrino events at ICAL will 
significantly improve the determination of the neutrino mass hierarchy 
from this analysis \cite{CG}. 
Also, hadrons are the only signature of neutral current (NC)
events in the detector. While NC events are not affected by active
neutrino oscillations, they are sensitive to active--sterile
oscillations; in addition, the NC rates are altered by the contribution
of tau-neutrinos (generated by active oscillations) through the hadronic
decay of taus produced in CC interactions. Hence, in order to understand
these events, it is important to characterise the behaviour of single
and multiple hadrons in the detector.

\begin{figure}[t]  
\centering\includegraphics[width=0.7\textwidth]{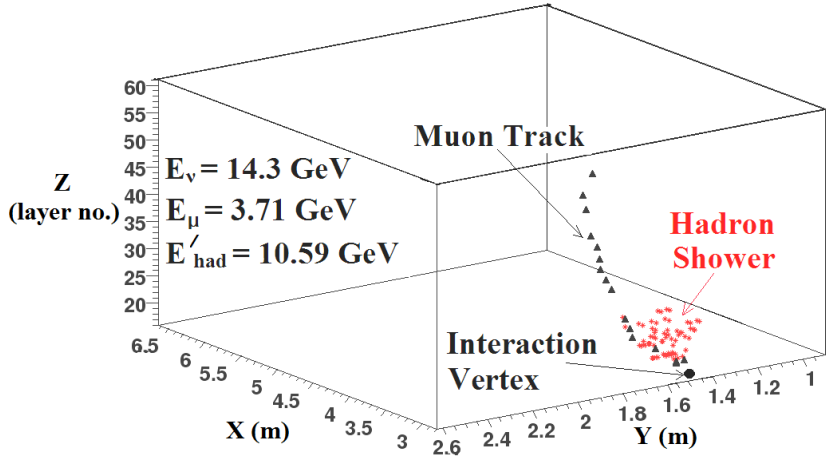}  
\caption{A deep inelastic muon-neutrino interaction event in the 
simulated  ICAL detector. Only the relevant part of the detector is 
shown. X and Y indicate length in units of meters whereas Z 
indicates the layer index.} 
\label{visual}  
\end{figure}

The simulation study performed for estimating the hadron energy 
resolution for the ICAL detector is reported here. 
The hadrons consist mainly of pions  (both neutral and charged, 
and about 85\% of the events on the average), kaons, and also nucleons, 
including the recoil nucleon which cannot be distinguished from the 
remaining hadronic final state. The neutral pion 
decays immediately, giving rise to two photons, while the charged pions 
propagate and develop into a cascade due to strong interactions.  
A visualization of a neutrino DIS event with large 
hadron energy component in the simulated ICAL detector (using the VICE 
event display package \cite{vice}) is shown in Fig.~\ref{visual}. The 
main uncertainty in determining the incident neutrino energy comes from 
the uncertainty in estimating the energy of these hadrons. For 
the neutrino-nucleon interaction $\nu_{\mu} N \rightarrow \mu X$, 
the incident neutrino energy is given by 
\begin{equation} 
 \label{simple_equation} 
{\rm E}_{\nu} = {\rm E}_{\mu} + {\rm E}_{hadrons} - {\rm E}_N, 
\end{equation} 
where ${\rm E}_N$ is the energy of the initial nucleon which is taken 
to be at rest,  neglecting its small Fermi momentum. It can be seen from 
Eq.~(\ref{simple_equation}) that any uncertainty in measuring the hadron 
energy will directly affect the determination of incoming neutrino energy. 
Note that the visible hadron energy depends on factors like the shower 
energy fluctuation, leakage of energy, and invisible energy loss mechanisms, 
which in turn affect the energy resolution of hadrons . 
 
The analysis to obtain the hadron energy resolution is done in two parts.  
First, the charged pions of fixed energies are generated via Monte Carlo 
and propagated through the ICAL detector 
at fixed energies. This is used to calibrate the detector for 
the number of hits at fixed energies. Next, the multiple hadrons 
produced through neutrino interactions are considered, and the parameter 
$\textnormal{E}_{\textnormal{had}}^{\prime}$, defined as 
\begin{equation} 
 \label{simple_equation1} 
 \textnormal{E}_{\textnormal{had}}^{\prime} = \textnormal{E}_{\nu} -  
\textnormal{E}_{\mu}\; , 
\end{equation}  
is used to calibrate the detector response. Here, all types 
of hadrons contribute to the energy, though it is dominated by pions at 
the energies of a few GeV.

This paper describes the complete simulation chain of ICAL events 
which includes event generation through the event generator or 
particle gun, full detector simulation by propagating the events through 
the virtual detector and data analysis to assess the detector's capability 
to estimate hadron energy. In section~\ref{sec:E_res_fix},  
the hit distributions of charged hadrons, in particular, pions, in the 
ICAL detector are discussed. This is done by a Monte-Carlo (MC) generator 
that generates hadrons at fixed energies up to 15 GeV. 
The hits are then analysed for determining the detector response. 
The energy and its resolution are estimated using the hit multiplicity 
information of events generated with appropriate position- and 
angle-smearing. In section~\ref{sec:E_res_nuance}, the response of the
detector to multiple hadrons generated in atmospheric 
neutrino-nucleus interactions is studied using the neutrino event generator  
NUANCE \cite{Casper:2002sd} to obtain the net hadron energy resolution as a 
function of $\textnormal{E}_{\textnormal{had}}^{\prime}$.   
In section~\ref{sec:E_calib}, the calibration of 
$\textnormal{E}_{\textnormal{had}}^{\prime}$ using the detector hit 
information is described. 
Our results are summarized in section~\ref{sec:concl}.

\section{Energy response to fixed-energy hadrons} 
\label{sec:E_res_fix} 

The ICAL detector is simulated using the GEANT4 package 
\cite{Agostinelli:2002hh,Brun:2000es}. The input parameters for 
the simulation -- like the gas mixture, conducting coating, efficiency, etc -- 
correspond to those for a 12-layered stack of glass RPC detectors (without the 
Iron absorbers) \cite{Datar:2009zz} that is being run under stable conditions for several 
years. Iron layers, 5.6 cm thick, 
are interleaved with the active RPCs with a 2 mm gas gap. The spacing 
between two consecutive RPCs is 9.6 cm. When a charged particle 
propagates through the ICAL detector, hits in the X and Y strips 
of the RPC layers are recorded. The layer number provides the $z$-coordinate. Thus 
the full position information is available, with a precision of 1.96 cm 
in the $x$ and $y$ directions and 2 mm in the $z$ direction. 
In this study we have neglected the hits due to edge effects and noise, and these will be 
taken care of once the data from the prototype detector becomes available in future. A check using the 
data \cite{mudirical} obtained from the cosmic muon study with the prototype
 mentioned above indicates that 
the hadron energy resolutions are not much affected by the edge effects. For more 
details about the GEANT4 based simulation of  
the detector geometry and the nature of hits 
generated, see Ref.~\cite{muonpaper}, which also discusses the muon energy  
response. The focus of this section will 
be the analysis of fixed-energy single-pion hit distributions.

\begin{figure}[h]  
\includegraphics[scale=0.39]{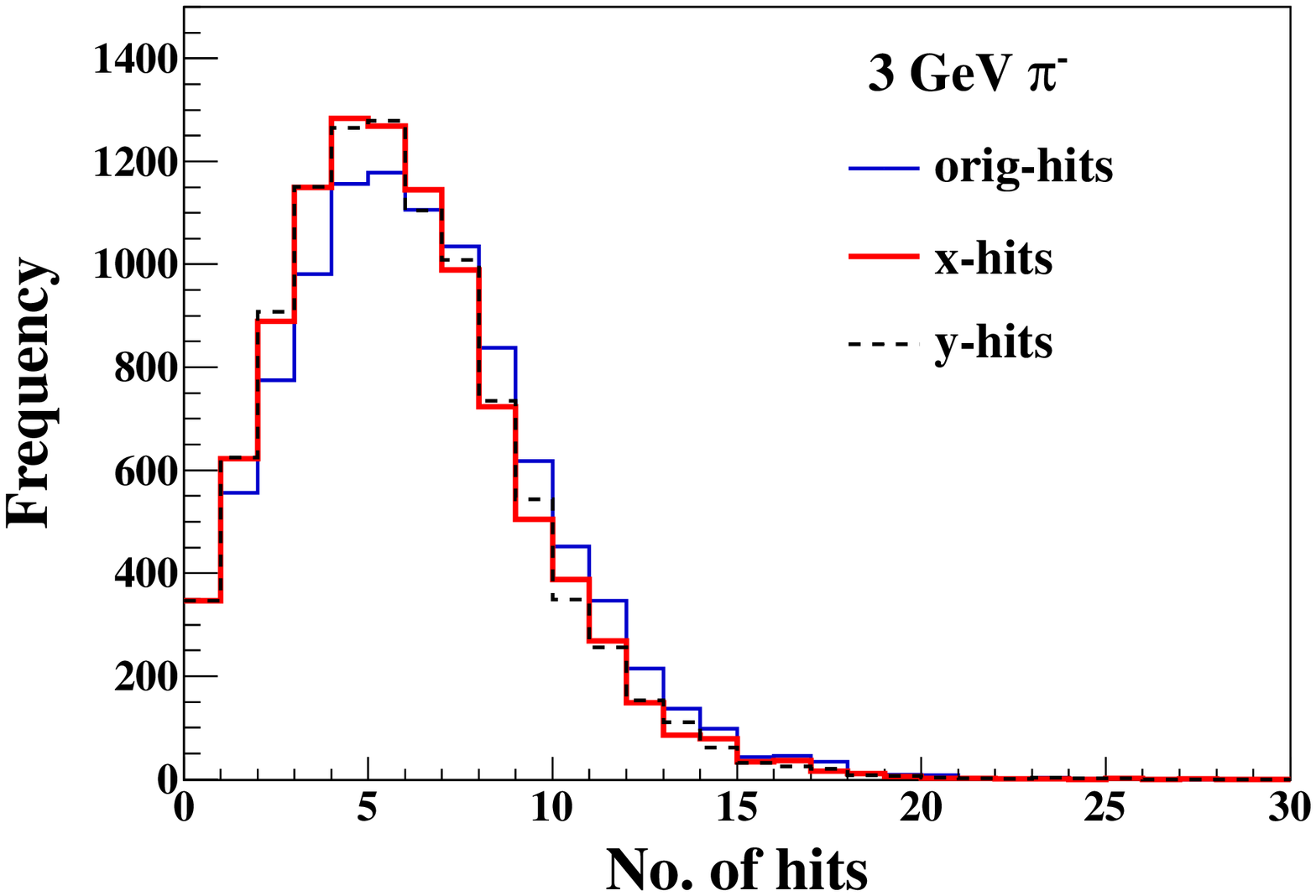}  
\includegraphics[scale=0.39]{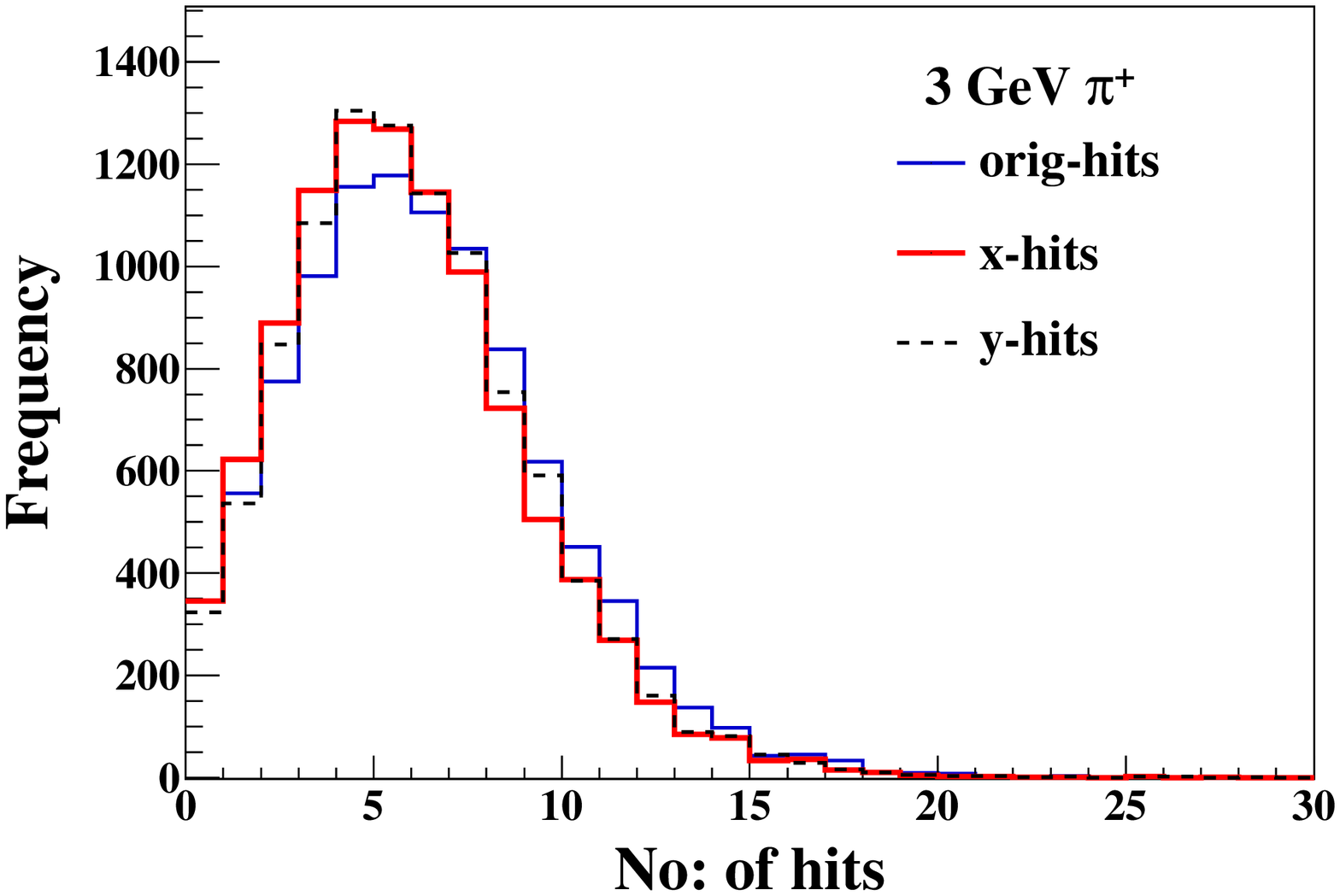} 
\caption{The comparison of the distributions of x-hits, y-hits and  
orig-hits for $\pi^{-}$ (left) and $\pi^{+}$ (right) of energy 3 GeV.}  
\label{xyorcomp}  
\end{figure}

A muon usually leaves one or two hits per layer and so the hits from both 
X and Y strips can easily be combined to obtain the number of hits and 
their position coordinates $(x, y)$ in a given layer. However, in the 
case of hadron shower there are multiple hits per layer, and combining X 
and Y strip hits leads to some false count of hits (ghost hits). To avoid 
the ghost hit counts, the energy calibration may be done with counts from 
either X or Y strips. The variables x-hits and y-hits store the number of 
hits in the X and Y strips of the RPC, respectively. The maximum of x-hits 
or y-hits is stored as the variable ``orig-hits''. In Fig.~\ref{xyorcomp}, 
the comparisons of these three types of hit variables for 
$\pi^{\pm}$ of energy 3 GeV are shown. As is clear from Fig.~\ref{xyorcomp}, 
any of the variables x-, y- or orig-hits could have been used in the analysis. 
The variable orig-hits has been chosen as the unbiased parameter here. 
It is also observed that the detector response to the positively
and negatively charged pions is identical, so we shall not differentiate 
between them henceforth.

Fixed-energy single pion events in the energy range of 1 to 15~GeV 
were generated using the particle gun. Unless otherwise specified 
in this section, the total number of events 
generated for each input energy value is 10000. 
Each event is randomly generated 
over a volume 2 m $ \times $ 2 m $\times$ 2 m in the central 
region of the ICAL detector. In addition, since there is very little 
impact of the magnetic field on the showers produced by hadrons, the 
hadron direction is uniformly smeared over zenith angle $0 \le \theta 
\le \pi$ and azimuth of $ 0 \le \phi \le 2\pi$. (The angles are denoted 
with respect to a reference frame, where the origin is taken to be the 
center of the detector, the $z$-axis points vertically up, while the 
plates are horizontal in the $x$-$y$ plane.) This serves to smear out any 
angle-dependent bias in the energy resolution of the detector by virtue 
of its geometry which makes it the least (most) sensitive to particles 
propagating in the horizontal (vertical) direction.

\begin{figure}[ht]  
\includegraphics[scale=0.39]{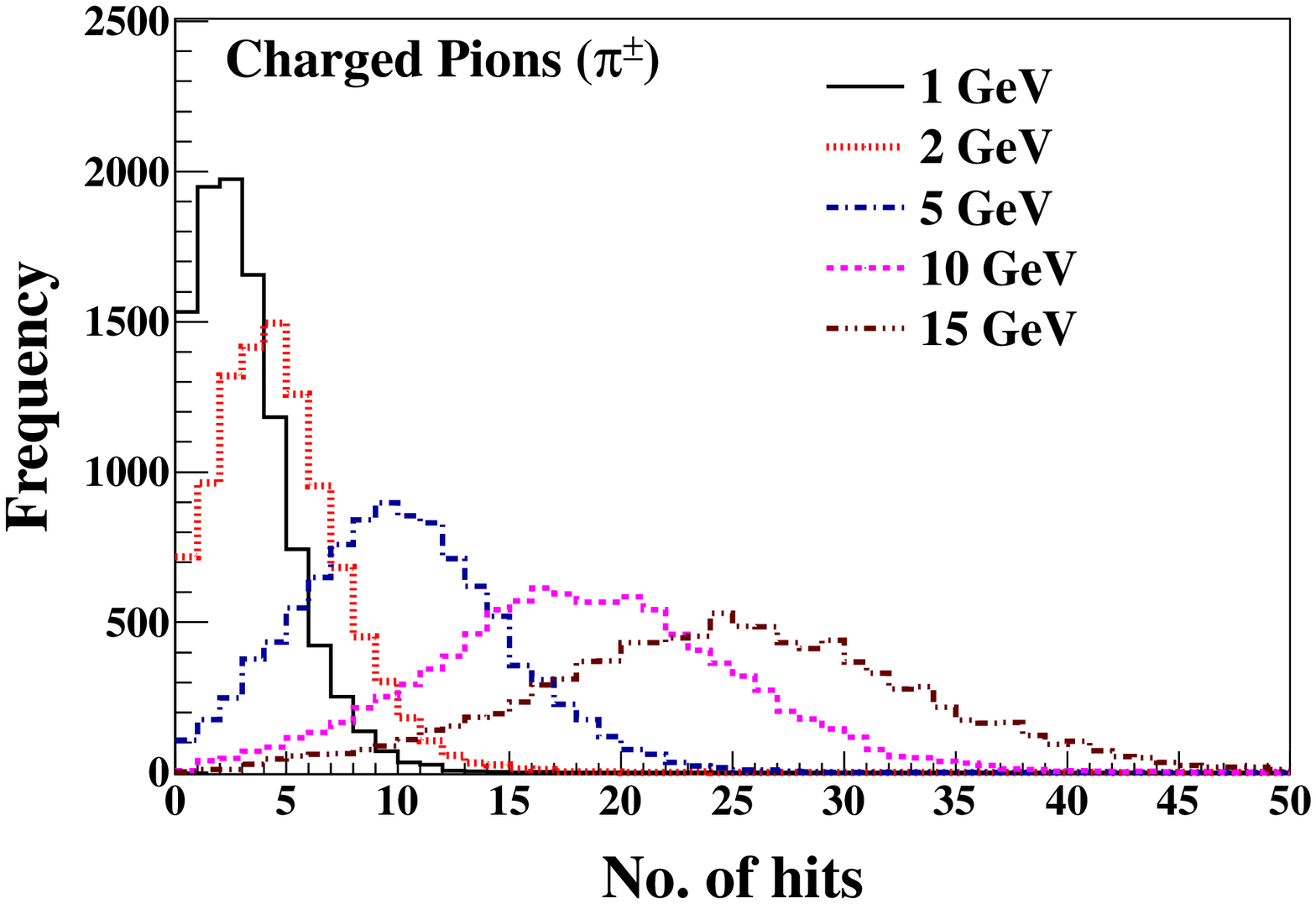}  
\includegraphics[scale=0.39]{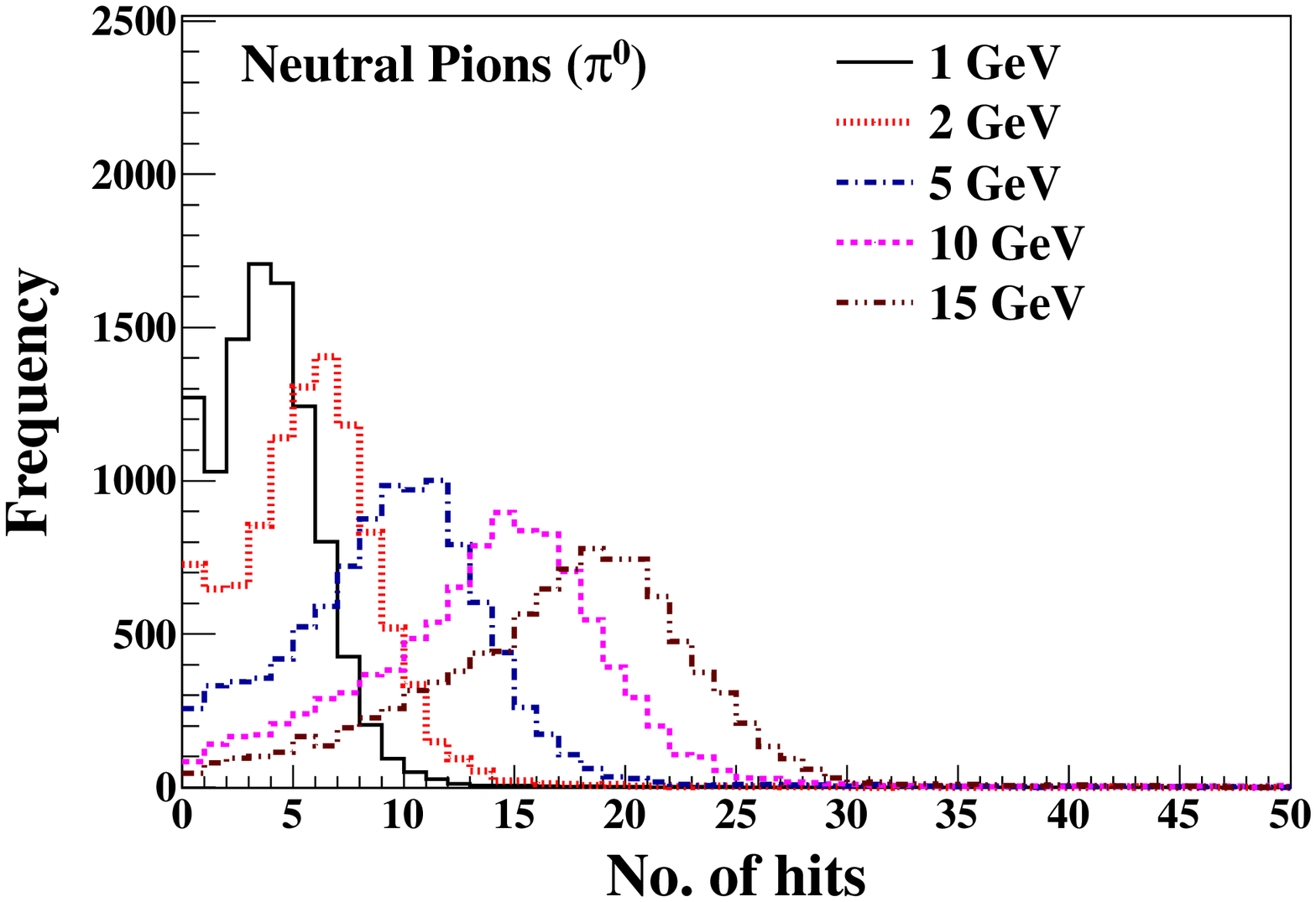}  
\includegraphics[scale=0.39]{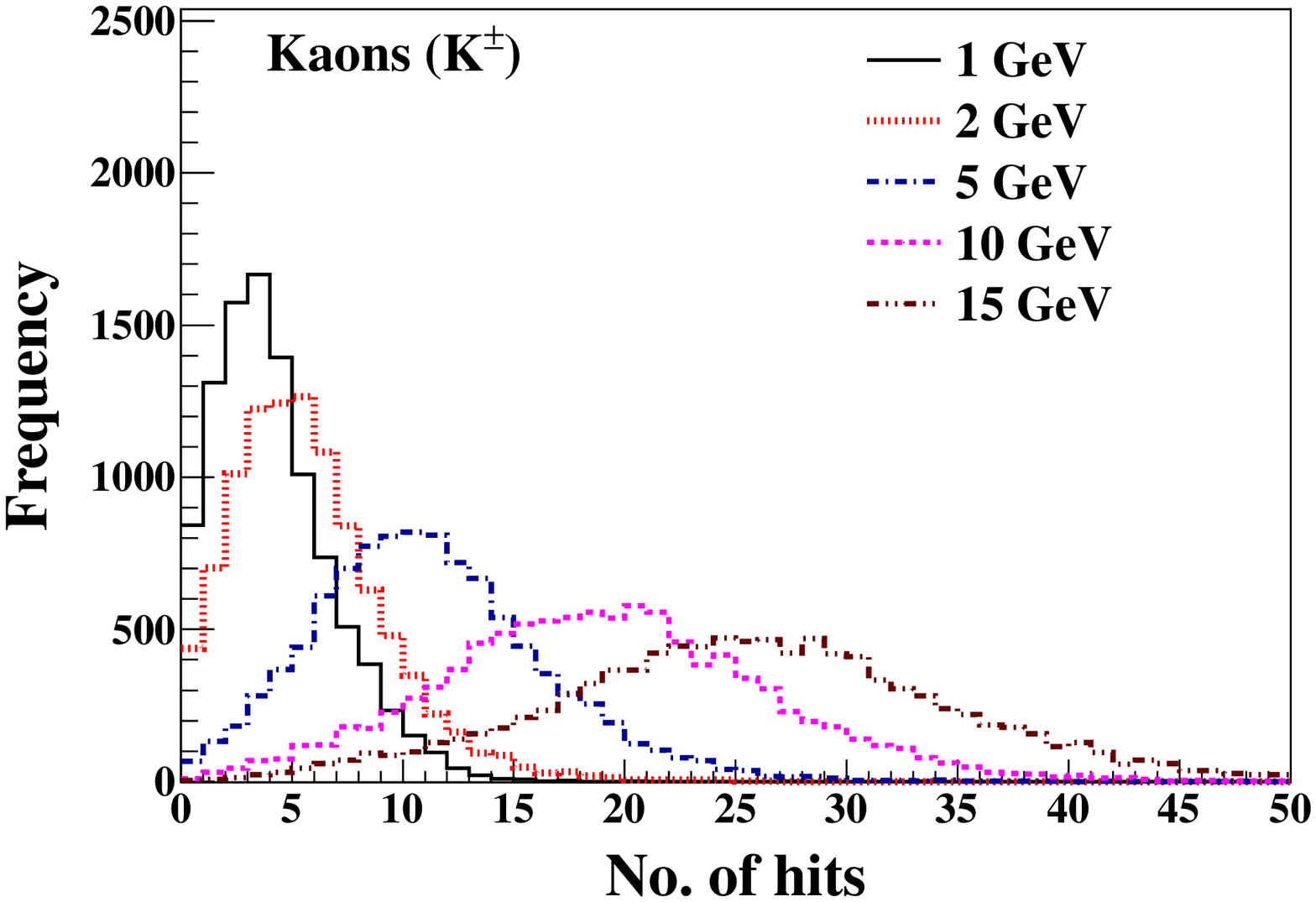}  
\includegraphics[scale=0.39]{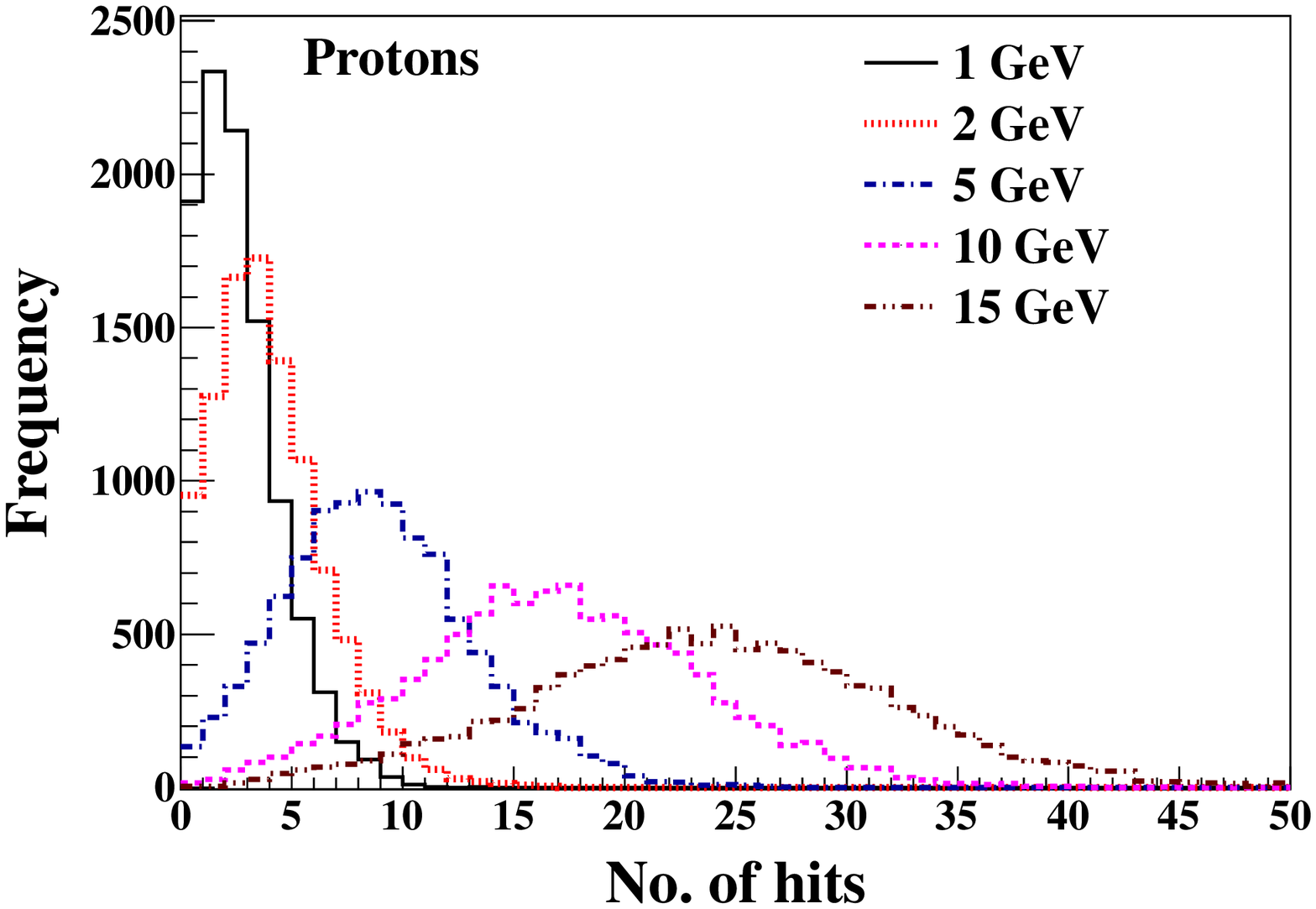}  
\caption{The hit distributions at various energies (angle-averaged) 
for $\pi^{\pm}$, $\pi^{0}$, $K^{\pm}$ and protons propagated from 
vertices smeared  over the chosen detector volume.} 
\label{histocomp}  
\end{figure}

In Fig.~\ref{histocomp}, the hit distributions in the detector for 
pions, kaons, and protons at various energies in the range of 1 to 15 
GeV are shown.  It is observed that for all these hadrons the hit patterns 
are similar, though the peak position and spread are somewhat dependent 
on the particle ID. Hence the detector cannot distinguish the specific 
hadron that has generated the shower. The large variation in the number 
of hits for the same incident particle energy is partly due to angle 
smearing and more dominantly due to the strong interaction processes 
with which hadrons interact with the detector elements. The exception is 
$\pi^0$, which decays almost immediately into an $e^+\,e^-$ pair; the 
fewer number of hits and the narrower hit distribution in this case 
reflects the nature of electromagnetic interactions of this pair with the 
predominantly iron target. 

\newpage
Since hadrons produced in neutrino interactions with ICAL are primarily 
charged pions, the focus in this section is on the detector response to 
charged pions. A more general admixture of different 
hadrons is considered in the next section.

\subsection{Analysis of the charged pion hit pattern} 
\label{sec:charge_pions} 

\begin{figure}[ht] 
\includegraphics[width=7.8cm] 
{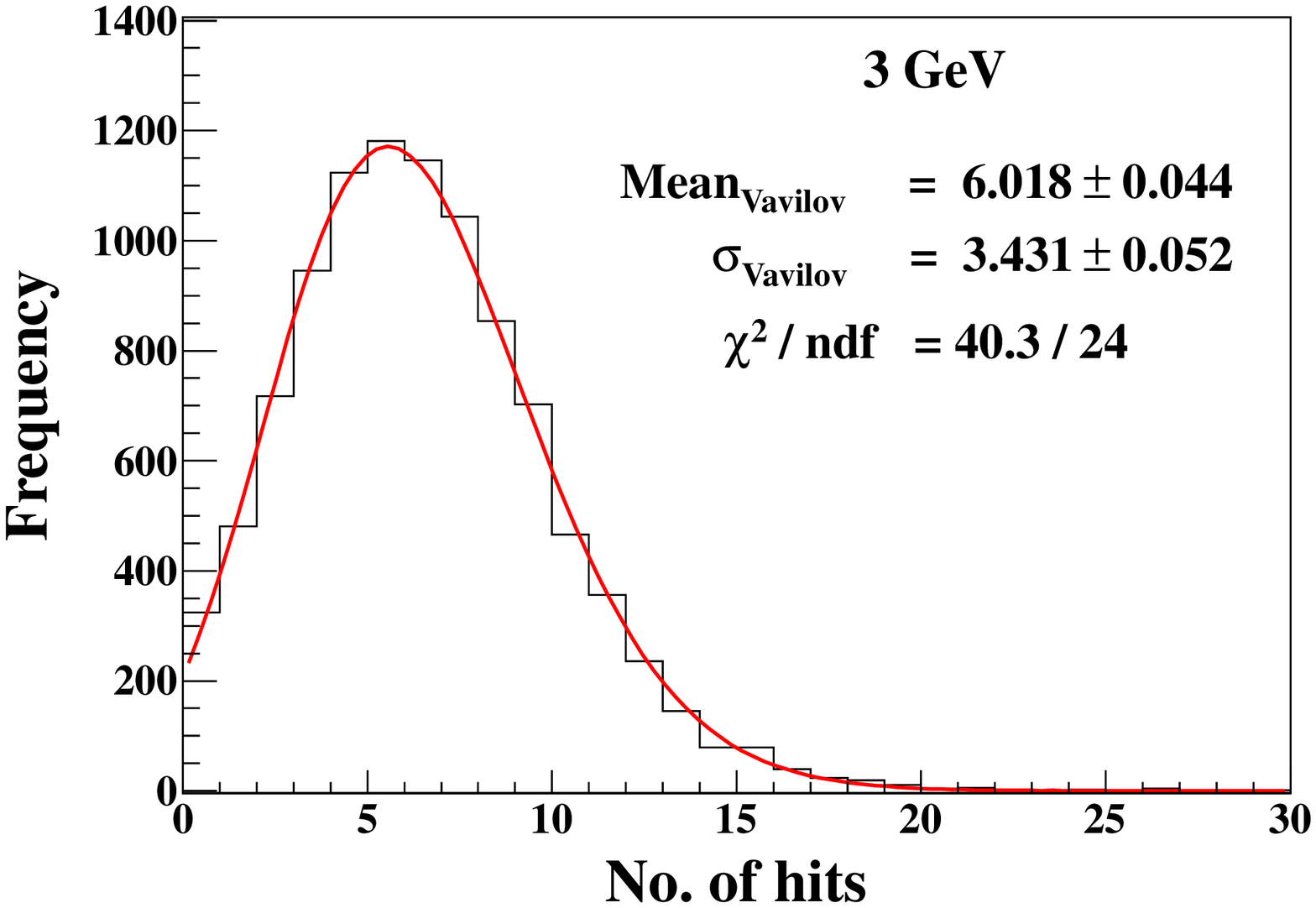}  
\includegraphics[width=7.8cm] 
{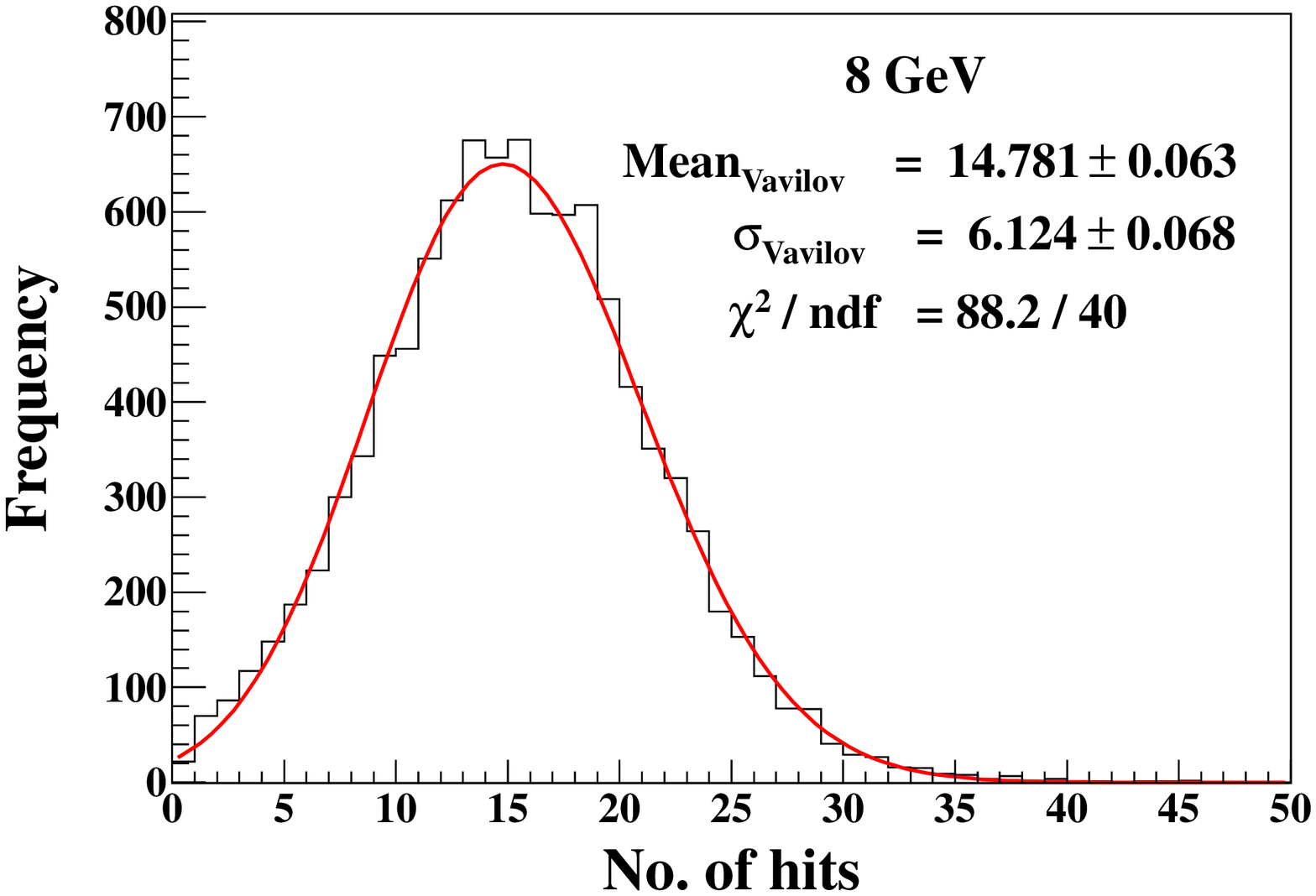} 
\caption{The hit distributions at 3 GeV (left) and 8 
GeV (right), for pions propagating in the detector, starting from randomized 
vertices over a volume of 2 m$~\times~$2 m$~\times~$2 m in the detector. 
The red curve denotes a fit to the Vavilov distribution} 
\label{pionhisto} 
\end{figure}

The charged pion hit distributions at sample values of $E=3,8$ GeV are 
shown in Fig.~\ref{pionhisto}. Typical patterns show a mean 
of roughly 2 hits per GeV as seen from the figure, but with long tails, 
so the distribution is not symmetric. In addition, several events yield 
zero hits in the detector at lower energies; such events are virtually 
absent at higher energies. 

\begin{figure}[t] 
\includegraphics[width=7.8cm]{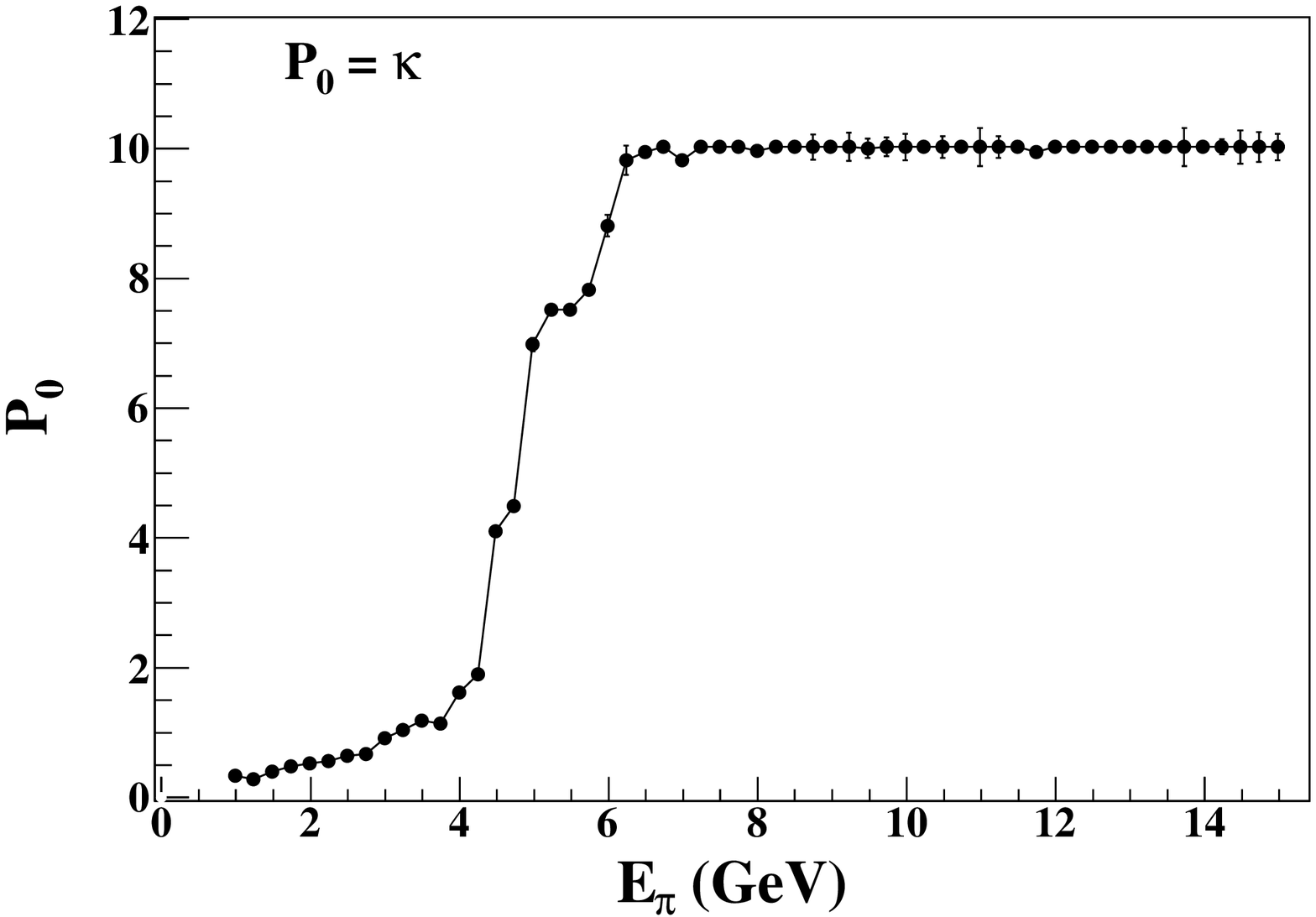}  
\includegraphics[width=7.8cm]{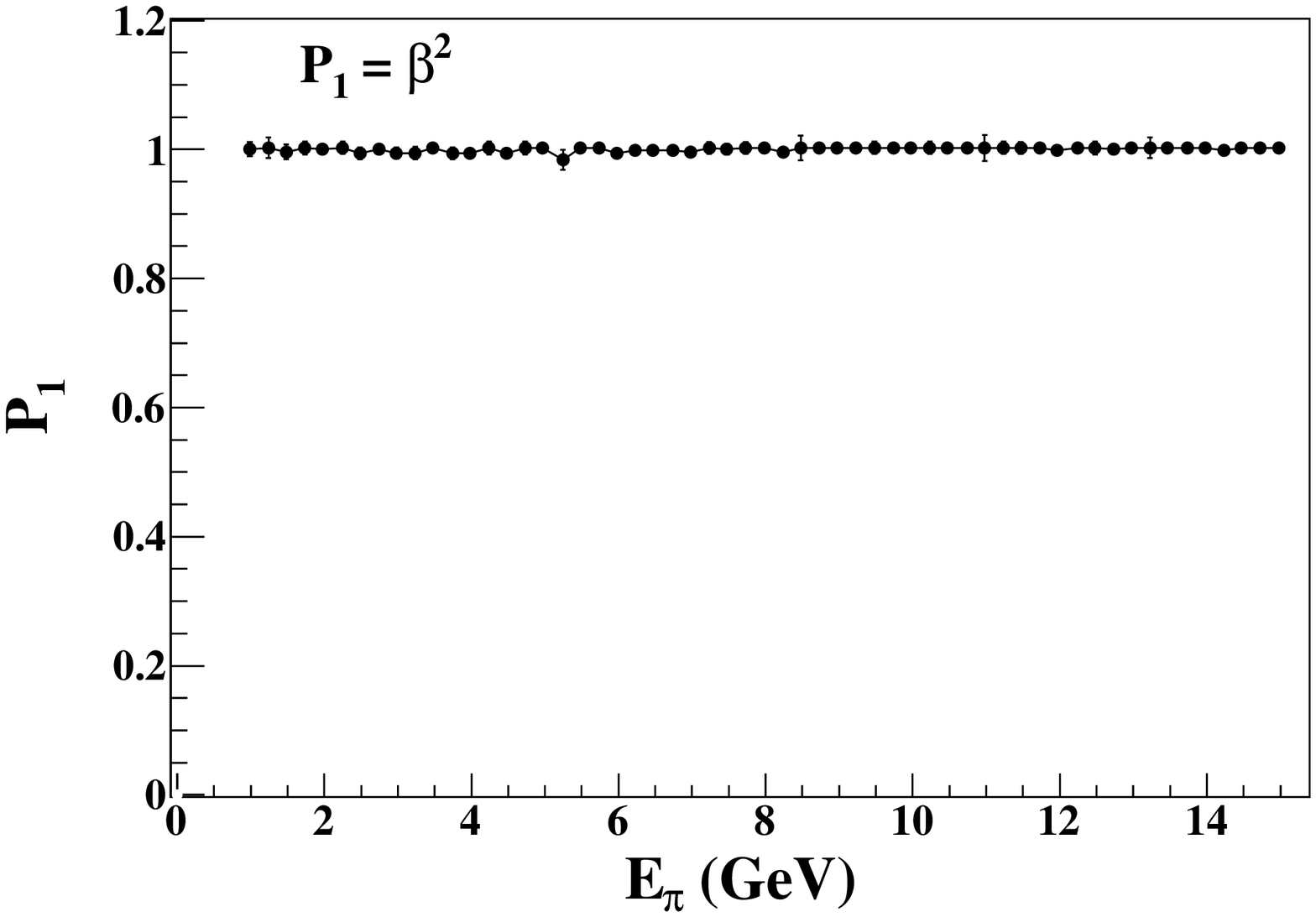}
\includegraphics[width=7.8cm]{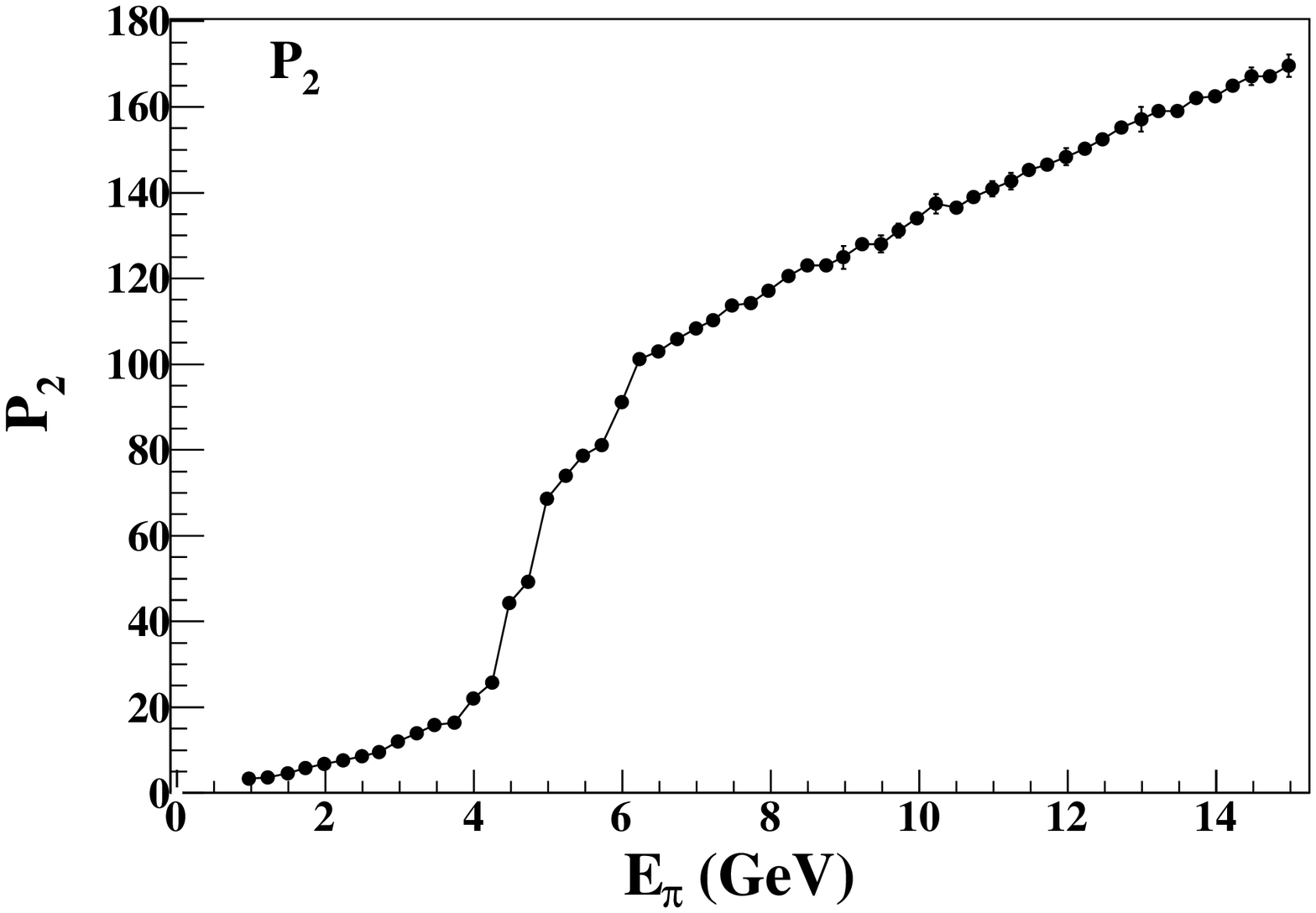}
\includegraphics[width=7.8cm]{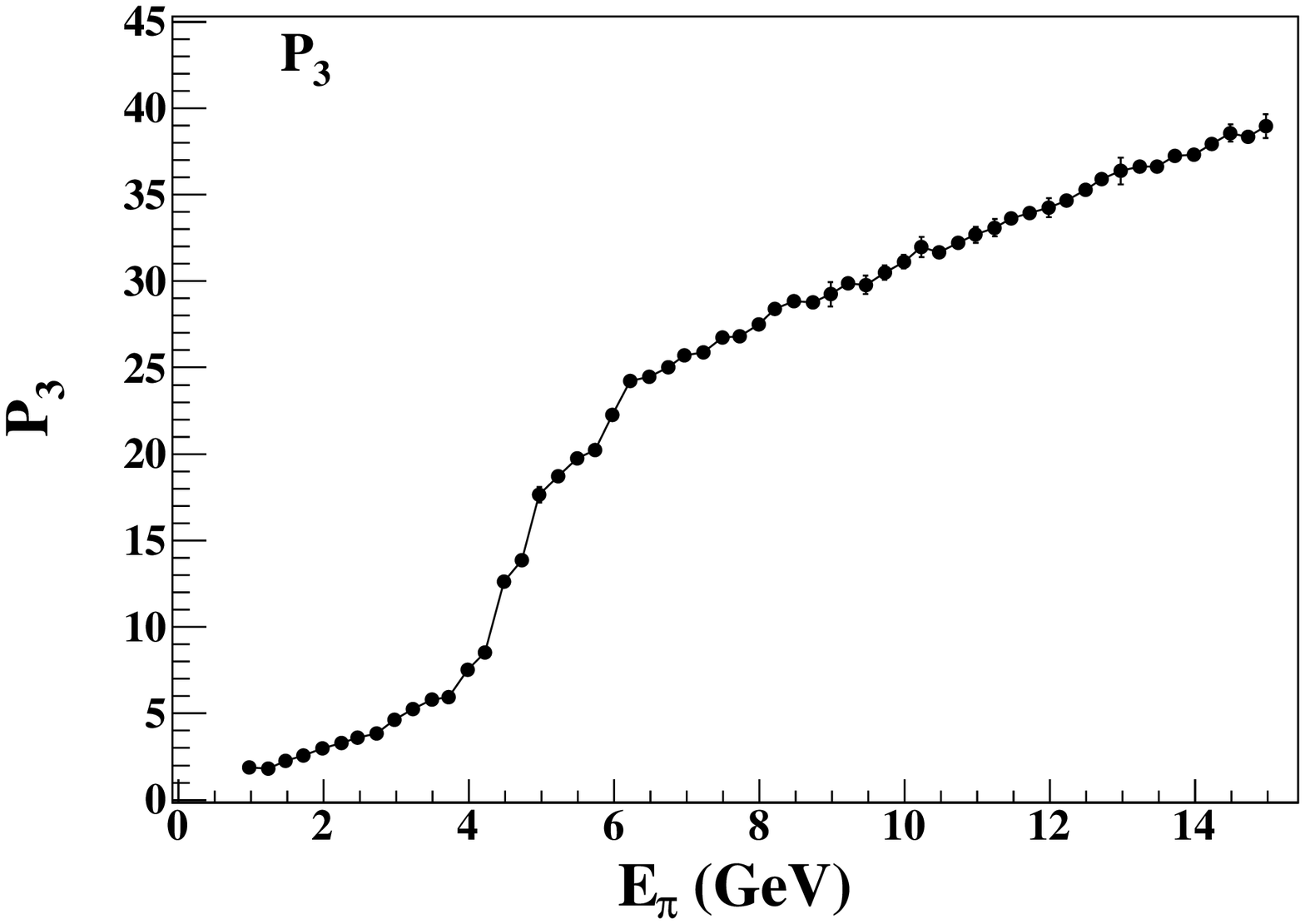} 
\caption{The parameters $\rm{P}_0$, $\rm{P}_1$, $\rm{P}_2$ and 
$\rm{P}_3$ of the Vavilov fit to the hit multiplicity,
as functions of pion energy, for fixed-energy charged pions.} 
\label{mcvavpar} 
\end{figure}

A good fit is obtained with the Vavilov distribution function 
for all energies, as is illustrated in Fig.~\ref{pionhisto}. 
This distribution (see Appendix \ref{sec:appA}) is described by 
the four parameters $\rm{P}_0$, $\rm{P}_1$, $\rm{P}_2$ and $\rm{P}_3$. 
The energy dependence of these parameters is shown in Fig.~\ref{mcvavpar}. 
This may be used directly for reconstructing the hit distribution at any 
given energy. Note that the Vavilov distribution reduces to a Gaussian 
distribution for $\rm{P}_0~\geq~10$. In this analysis it is observed 
to happen at energies greater than 6 GeV, as can be seen from 
Fig.~\ref{mcvavpar}. At lower energies, it is necessary to use the 
full Vavilov distribution.

\begin{figure}[h] 
\includegraphics[width=7.8cm]{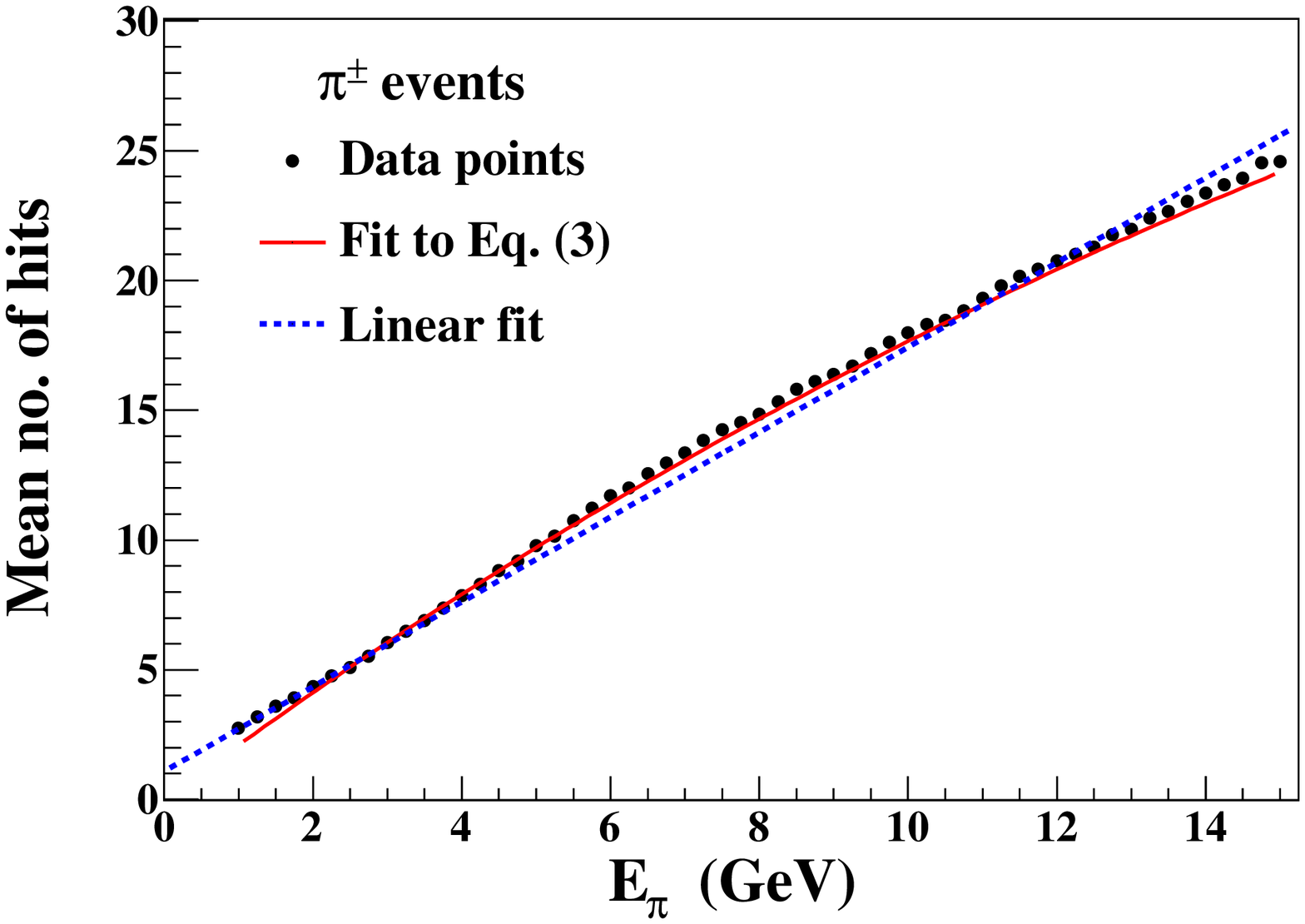}  
\includegraphics[width=7.8cm]{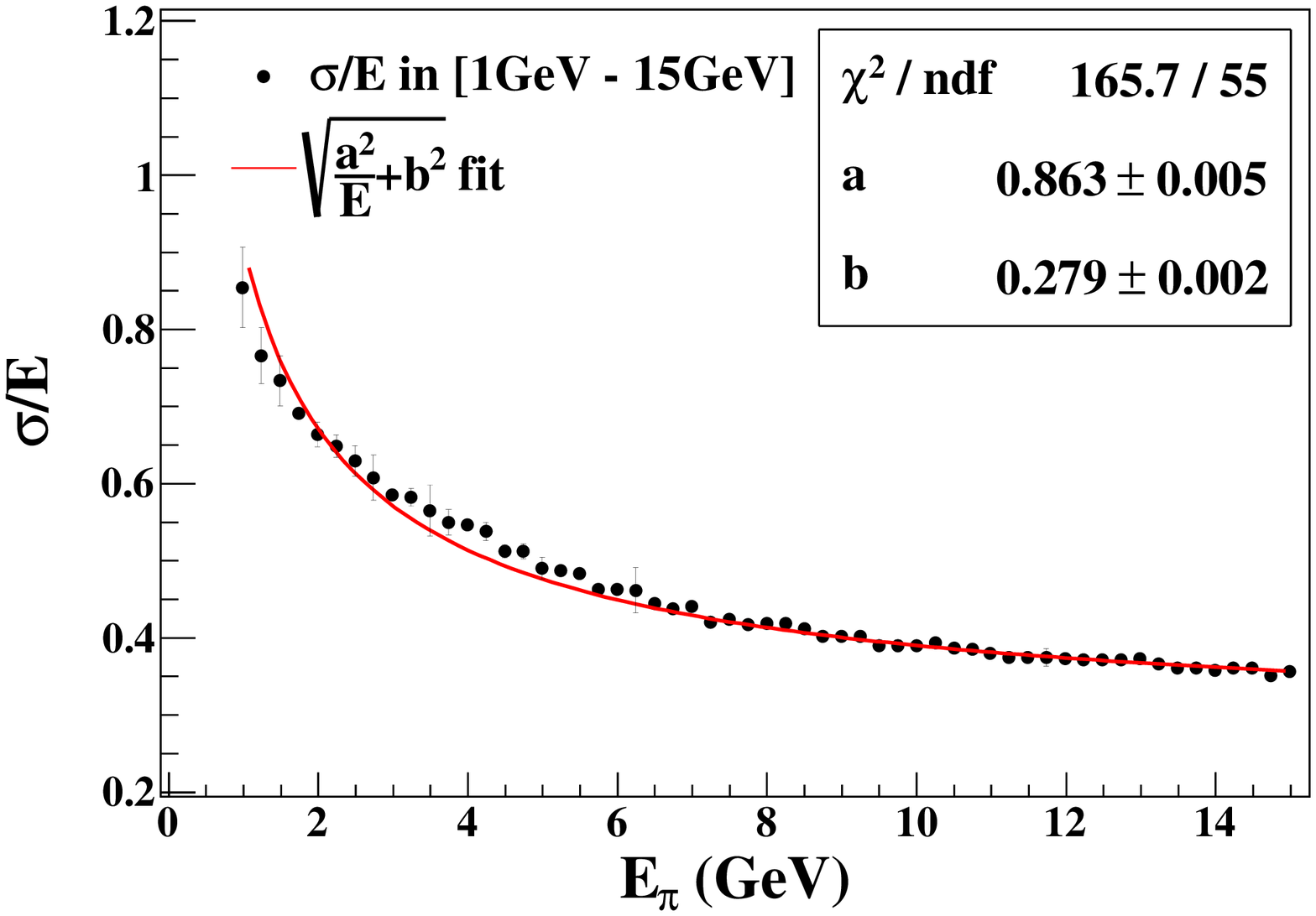} 
\caption{The mean hit distribution (left) and the energy resolution (right) 
for fixed-energy charged pion events, as a function of pion energy. 
The right panel also shows a fit to Eq.~(\ref{eq5}).} 
\label{meancomp} 
\end{figure}

The mean $\bar{n}({\rm E})$ of the number of hits from the Vavilov fit 
at different energies is shown in the left panel of Fig.~\ref{meancomp}. 
It increases with increasing pion energy, and saturates at higher energies. 
It may be approximated by 
\begin{equation} 
\bar{n}({\rm E})=n_0[1-\exp(-{\rm E}/{\rm E}_0)] \; , 
\label{eq2} 
\end{equation} 
where $n_0$ and ${\rm E}_0$ are constants. This fit has to be interpreted 
with some care, since $n_0$ and ${\rm E}_0$ are sensitive to the energy 
ranges of the fit. The value of ${\rm E}_0$ is found to be $\sim$ 30 GeV
when a fit to the energy range 1--15 GeV is performed.
Since the energies of interest for atmospheric neutrinos 
are much less than ${\rm E}_0$, Eq.~(\ref{eq2}) may be used in its approximate 
linear form $\bar{n}({\rm E})=n_0 {\rm E}/{\rm E}_0$. 
A fit to this linear form is also shown in Fig.~\ref{meancomp}.

Since in the linear regime (${\rm E} \ll {\rm E}_0$) one has
\begin{equation} 
\frac{\bar{n}({\rm E})}{n_0}=\frac{{\rm E}}{{\rm E}_0} \; , 
\label{eq3} 
\end{equation} 
The energy resolution may be written as 
\begin{equation} 
\frac{\sigma}{{\rm E}} 
 = \frac{\Delta {n}({\rm E})}{\bar{n}({\rm E})} \; , 
\label{eq4} 
\end{equation} 
where $(\Delta n)^2$ is the variance of the distribution.
In the rest of the paper  the notation $\sigma/{\rm E}$ will be used 
for energy resolution, and Eq.~(\ref{eq4}) 
will be taken to be valid for the rest of the analysis. 
 
The energy resolution of pions may be parameterized by  
\begin{equation} 
\frac{\sigma}{{\rm E}} = \sqrt{\frac{\rm{a}^2}{{\rm E}} + {\rm b}^2}~, 
\label{eq5} 
\end{equation} 
where ${\rm a}$ and ${\rm b}$ are constants.  
The energy resolutions for charged pions functions of pion energy are shown in  
Fig.~\ref{meancomp}. 
The parameters ${\rm a}$ and ${\rm b}$ extracted by a fit to Eq.~(\ref{eq5}) 
over the pion energy range 1--15 GeV are shown in the
right panel of Fig.~\ref{meancomp}.
\newpage

\section{Energy response to hadrons produced by atmospheric neutrinos} 
\label{sec:E_res_nuance} 

The previous section contained an analysis of the energy resolution with 
single pion events. But in reality there are multiple 
hadrons produced in the atmospheric neutrino interactions. 
We analyze the charged-current (CC) $\nu_{\mu}$ interactions 
in the detector via quasi-elastic (QE), resonance, and deep inelastic 
scattering (DIS) processes. QE dominates at $E_\nu \sim 1$ 
GeV, and contains no hadron in the final state except for the recoil 
nucleon. Resonance events at a few GeV contain an additional hadron, 
typically a pion. As the energy increases, DIS events that contain  
multiple hadrons in the final state dominate.
 
Both atmospheric neutrino 
($\nu_\mu$) and anti-neutrino (${\overline{\nu}}_{\mu}$) events in 
ICAL are generated using the neutrino event generator NUANCE (v3.5) 
\cite{Casper:2002sd}. The hadrons produced in these interactions are  
primarily pions, but there are some 
events with kaons (about 10\%) and small fraction of other hadrons as well.  
As discussed earlier, it is not possible to discern one hadron from the 
other in the ``shower pattern'' of hadron hits. However, since the hit 
distribution of various hadrons are similar to each other 
(see Fig. \ref{histocomp}), 
and the NUANCE generator is expected to produce a correct 
mixture of different hadrons at all energies, it is sufficient to determine 
the hadron energy resolution at ICAL through an average of NUANCE events, 
without having to identify the hadrons separately. 
 
\begin{figure}[t] 
\includegraphics[width=7.8cm]{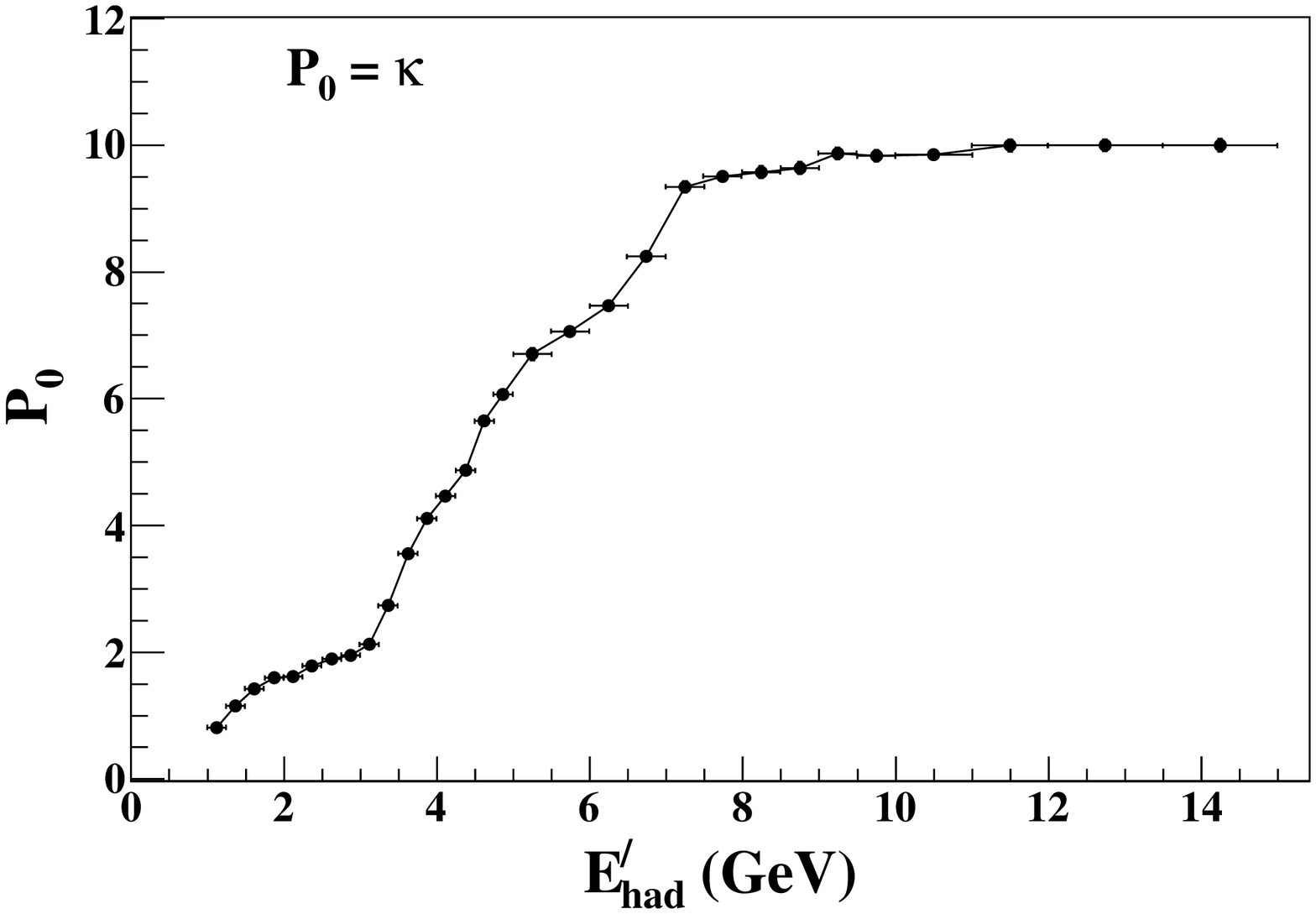} 
\includegraphics[width=7.8cm]{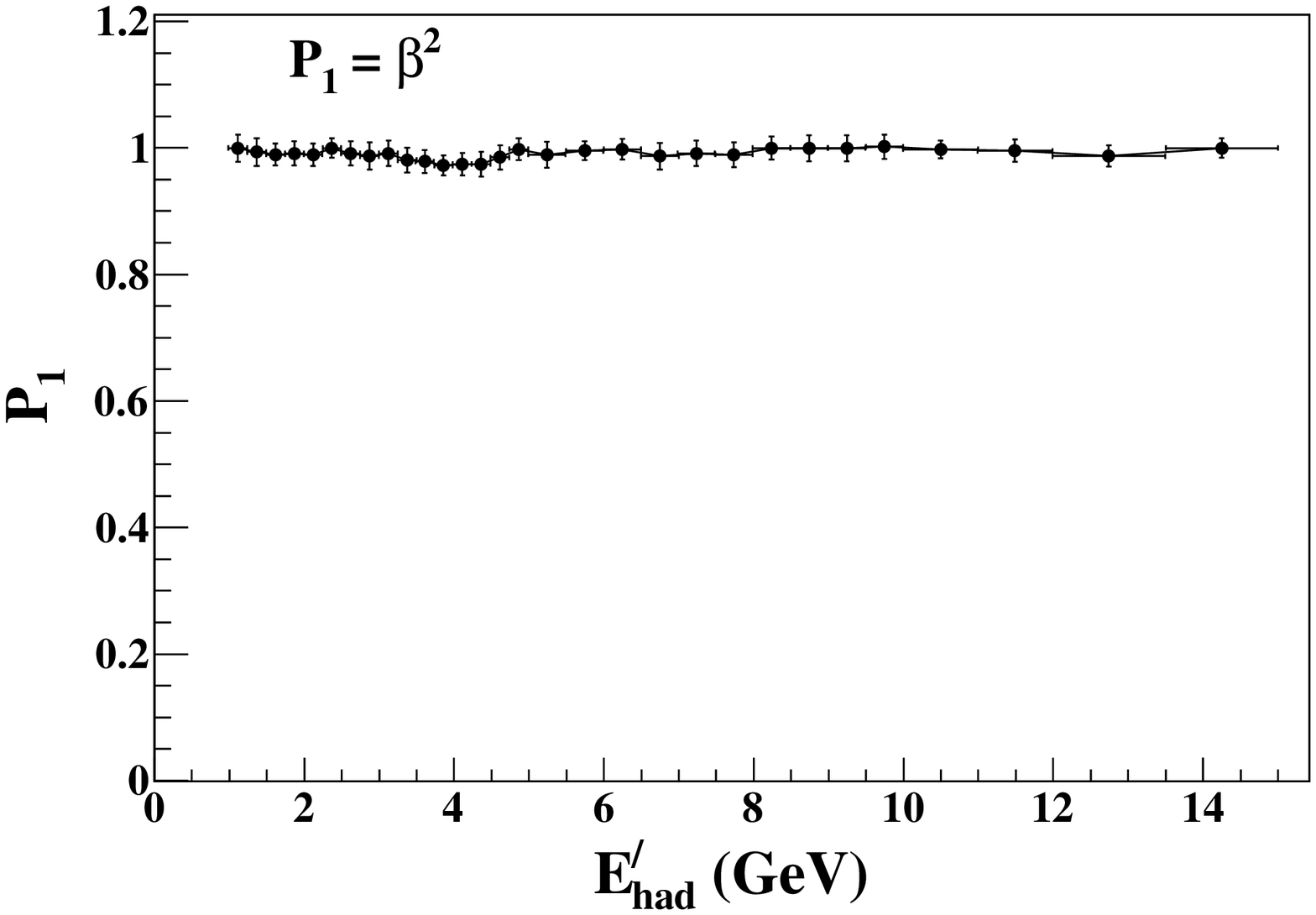}

\includegraphics[width=7.8cm]{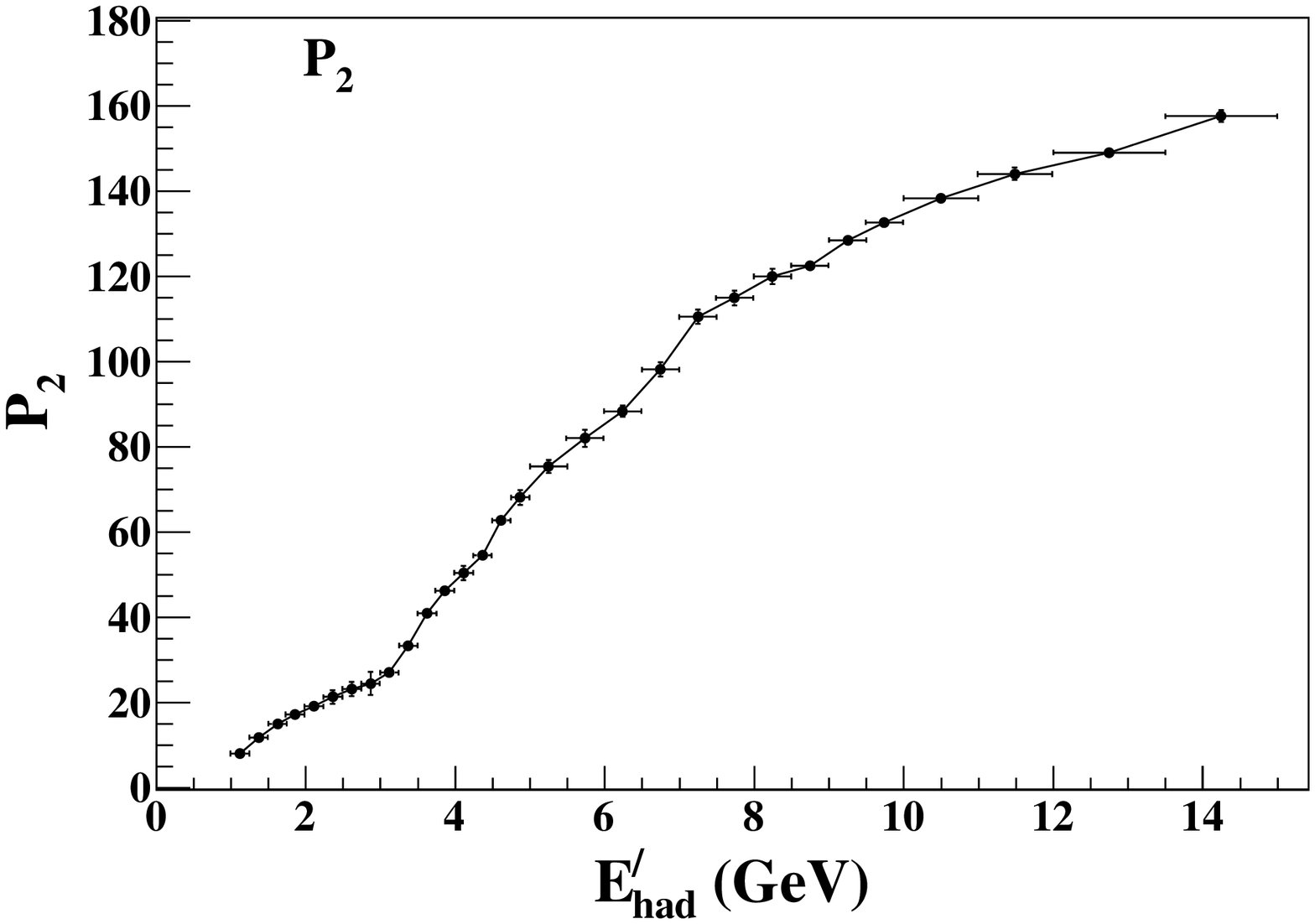} 
\includegraphics[width=7.8cm]{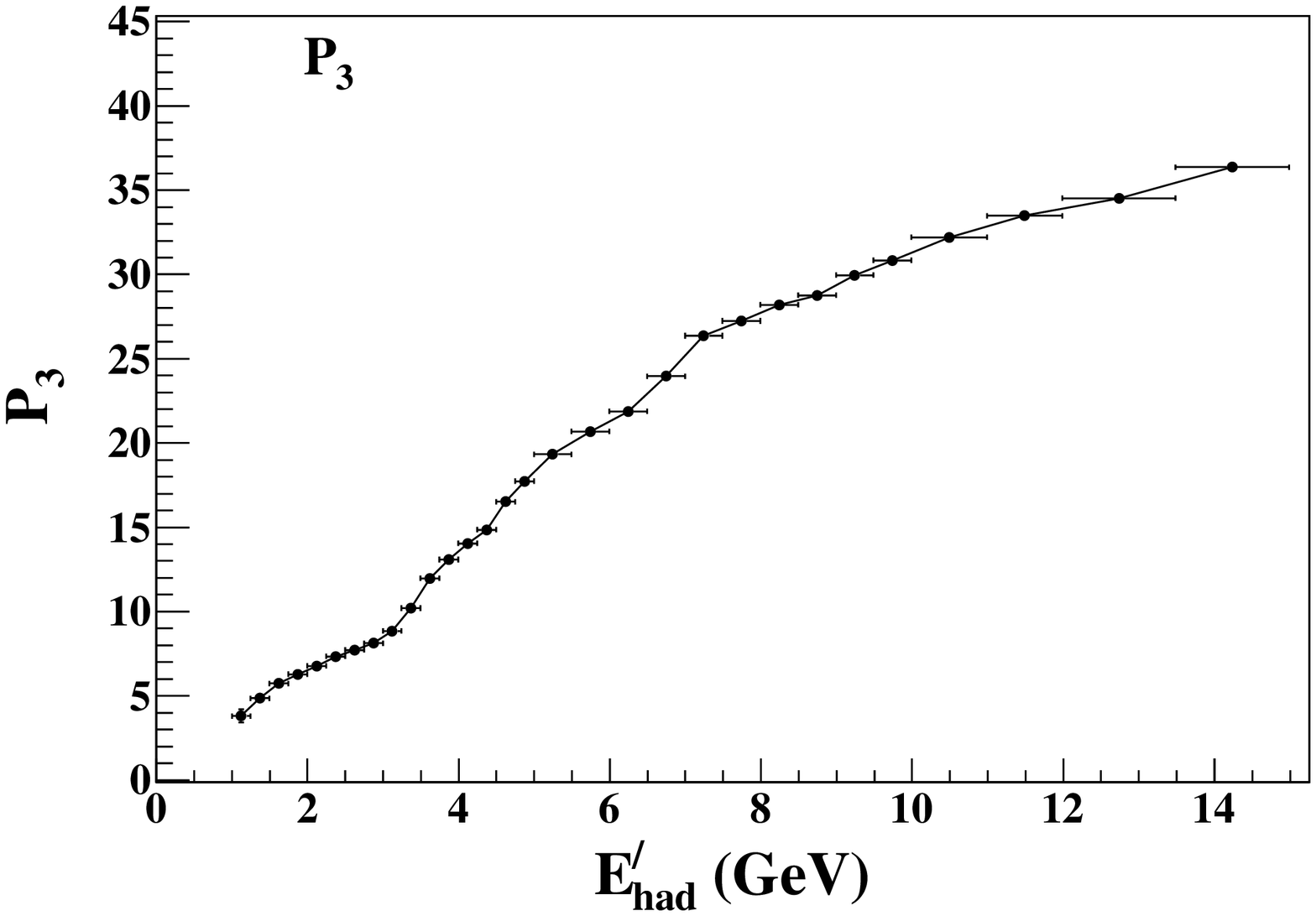}
\caption{The parameters $\rm{P}_0$, $\rm{P}_1$, $\rm{P}_2$ and 
$\rm{P}_3$ of the Vavilov fit to the hit multiplicity, 
as functions of $\rm{E}_{\rm{had}}^{\prime}$, from NUANCE data. 
These parameters can be directly used to reconstruct the hit distribution 
pattern. The bin widths are indicated by horizontal error bars. } 
\label{vav_par_nuance} 
\end{figure}

\begin{figure}[t] 
\includegraphics[width=7.8cm]{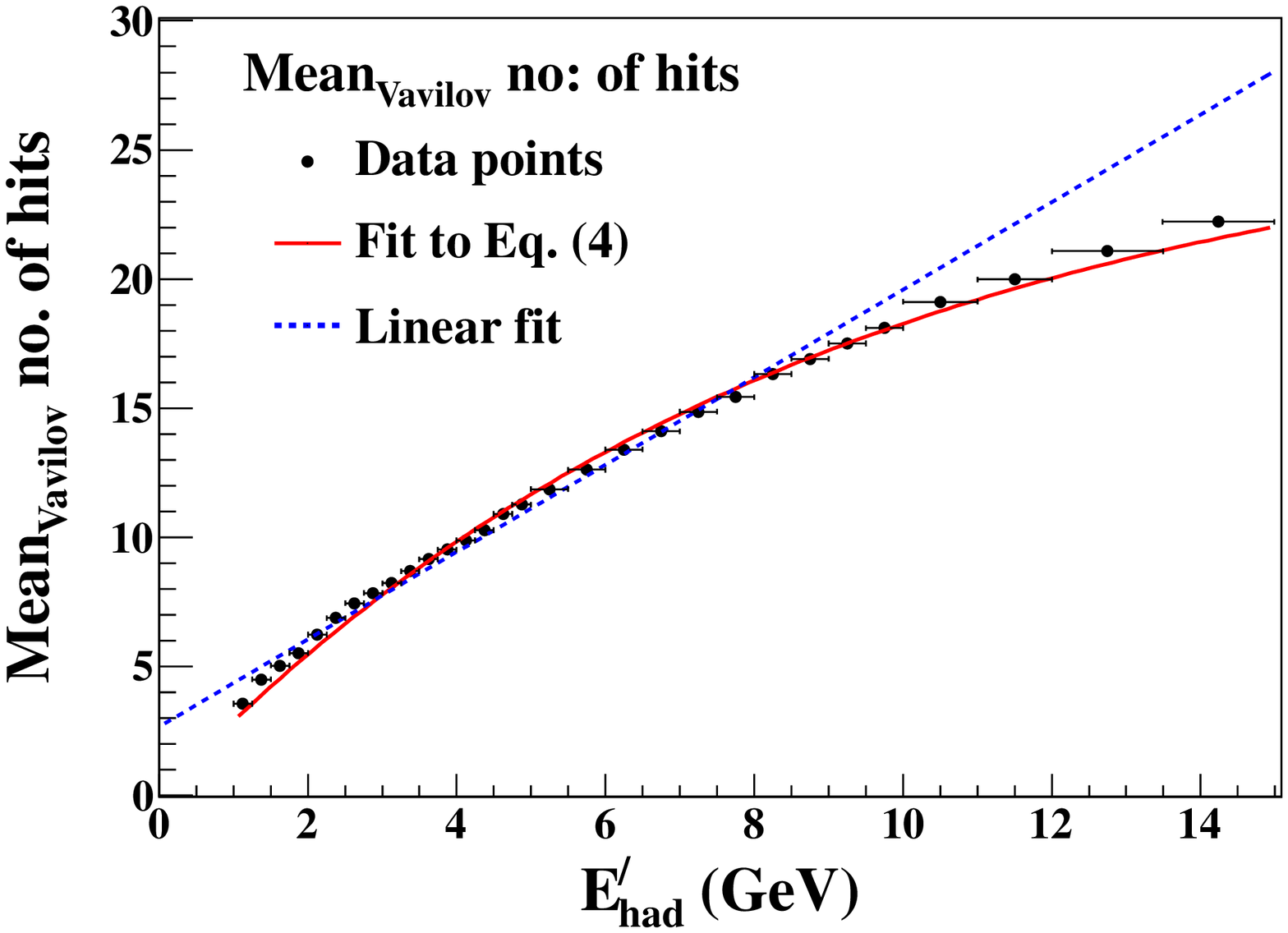} 
\includegraphics[width=7.8cm]{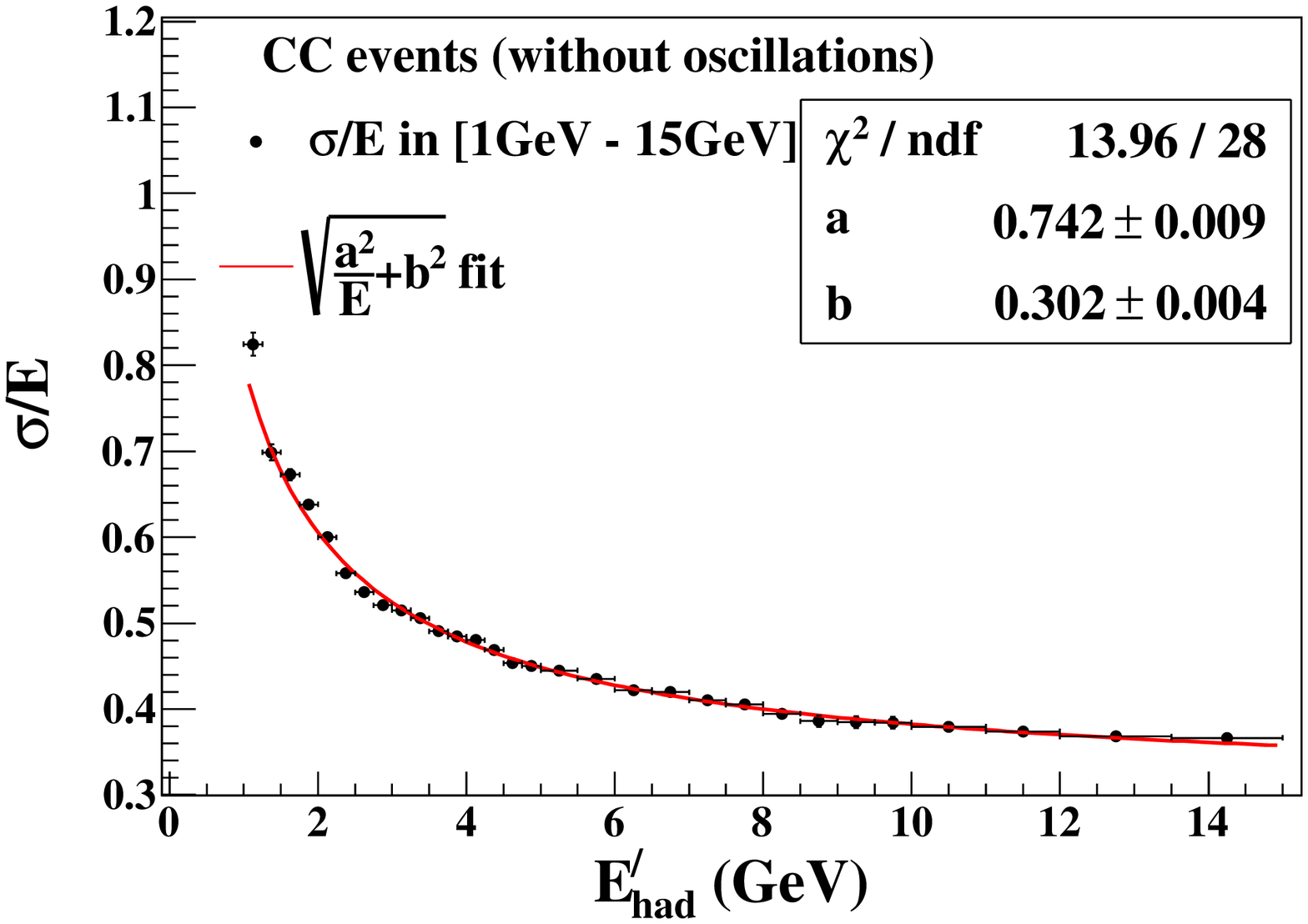} 
\caption{The mean hit distribution (left) and the energy resolution (right) 
for hadron events generated by NUANCE, as a function of 
$\rm{E}_{\rm{had}}^{\prime}$ . The right panel 
also shows a fit to Eq.~(\ref{eq5}). 
The bin widths are indicated by horizontal error bars.} 
\label{histomean_vav_nuance} 
\end{figure}

A total of 1000 kt-years of ``data'' events (equivalent to 20 
years of exposure with the 50 kton ICAL module) were generated with NUANCE. 
The events were further binned into the various 
$\textnormal{E}^{\prime}_{\textnormal{had}}$ energy bins and the hit 
distributions (averaged over all angles) in these bins are fitted to 
the Vavilov distribution function. 
The energy dependence of the Vavilov fit parameters is shown in 
Fig.~\ref{vav_par_nuance}. This information
can be used directly to simulate the hadron energy response of the 
detector for physics analysis.
The mean values ($\textnormal{Mean}_{\textnormal{Vavilov}}$) of  
these distributions as a function of ${\rm E}_{\rm{had}}^{\prime}$ 
are shown in the left panel of Fig.~\ref{histomean_vav_nuance}. 
As expected, these are similar to the mean values obtained earlier with fixed 
energy pions. Since the mean hits grow approximately linearly with energy, 
the same linearized approximation used in section~\ref{sec:charge_pions} 
can be used to obtain the energy resolution 
$\sigma/{\rm E}$ = $\Delta n/\bar{n}$. 
The energy resolution as a function of $\rm{E}_{\rm{had}}^{\prime}$ 
is shown in Fig.~\ref{histomean_vav_nuance}. 
The energy resolution ranges from 85\% (at 1 GeV) to 36\% (at 15 GeV).
\newpage 
The effective energy response obtained from the 
NUANCE-generated data is an average over the mixture of many hadrons 
that contribute to hadron shower at all energies. The fractional weights 
of different kinds of hadrons produced in neutrino interactions may, in 
principle, depend upon neutrino oscillations. In addition, the relative 
weights of events with different energy that contribute in a single energy 
bin changes because neutrino oscillations are energy dependent.  
Events with oscillations using  
the best-fit values of standard oscillation parameters  
(mixing angles and mass-squared differences) \cite{oscparam} 
were also generated. The resolutions obtained without and with 
oscillations are very close to each other. Thus, the hadron energy 
resolution can be taken to be insensitive to oscillations.

\section{Hadron energy calibration} 
\label{sec:E_calib} 

\begin{figure}[ht] 
\centering 
\includegraphics[width=7.8cm]{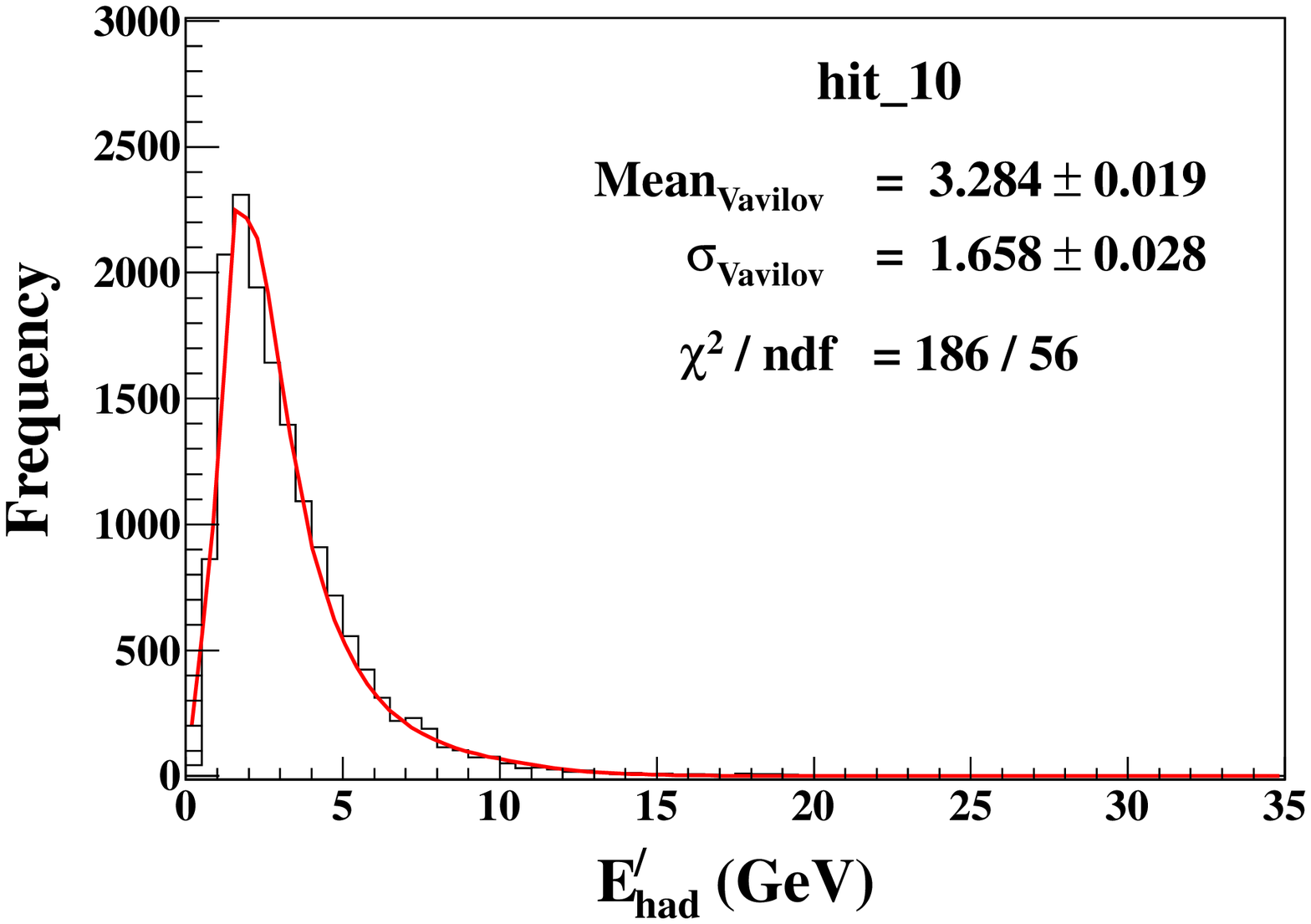} 
\includegraphics[width=7.8cm]{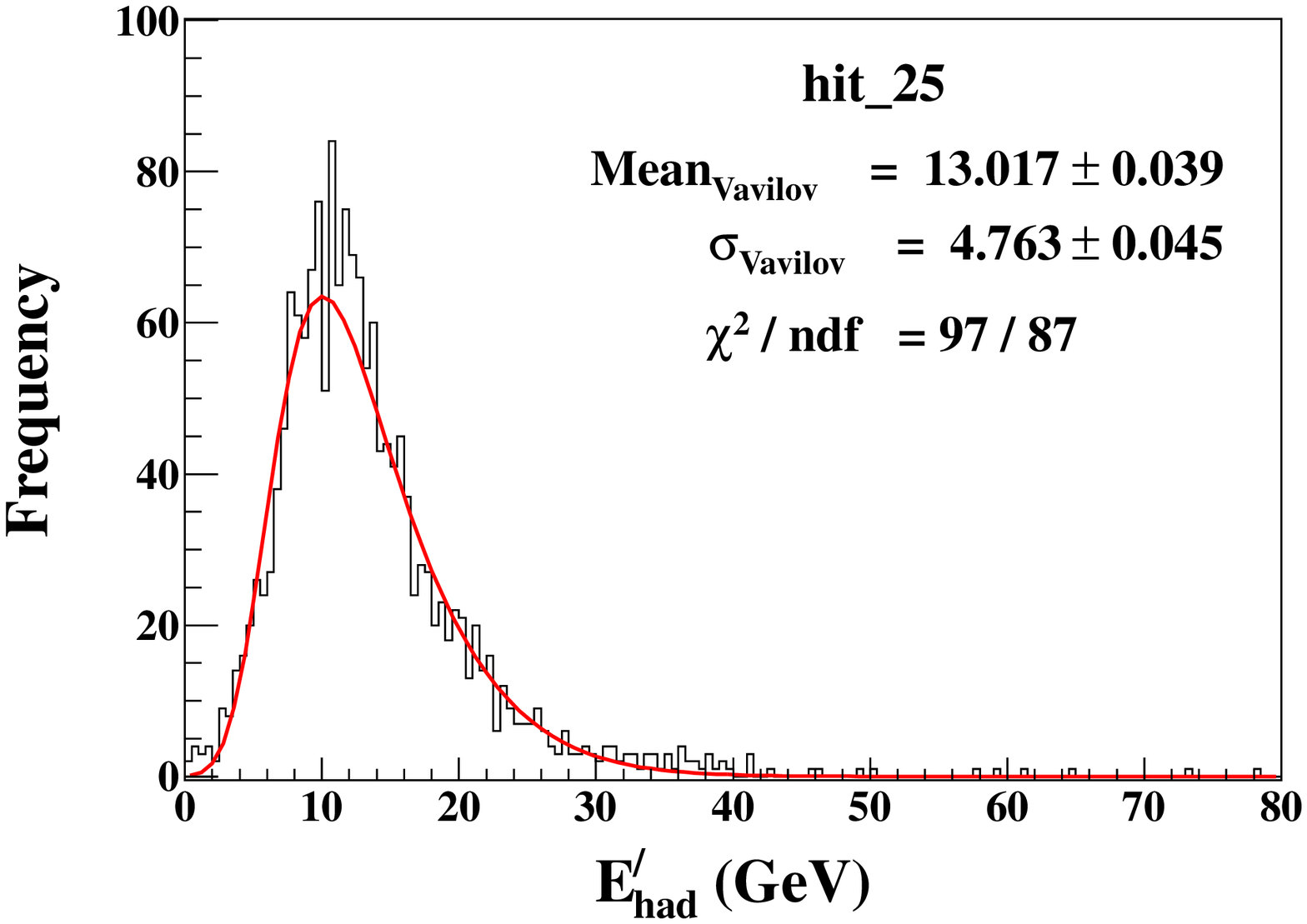} 
\caption{The distribution of hadron energy for a hit multiplicity 
of 10 (left) and 25 (right).}  
\label{calib_hit} 
\end{figure}

\begin{figure}[h] 
\centering 
\includegraphics[width=8cm]{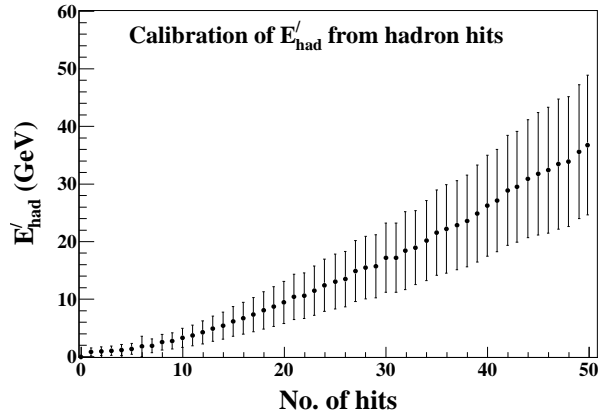} 
\caption{Calibration plot for $\textnormal{E}_{\textnormal{had}}^{\prime}$, 
where mean and $\sigma$ from the Vavilov fits are represented by the 
black filled circles and error bars, respectively.}  
\label{calib_nuance} 
\end{figure}

When the actual ICAL detector starts collecting the data, the only available 
observable for the hadrons is the hit multiplicity. Therefore, it is 
imperative  to calibrate this hit multiplicity with the hadron energy 
in the simulation. To this end, the hadrons from simulated NUANCE 
``data'' were divided into ``hit$\_\textnormal n$'' bins, where hit$\_$n 
corresponds to $\textnormal{n}$ number of hadron hits. The distributions of 
these energies were obtained for each bin as shown in Fig.~\ref{calib_hit}.
Even here, a good fit was obtained for the Vavilov distribution function 
at all hit multiplicities.The $\rm{Mean}_{\rm{Vavilov}}$ and $\sigma_{\rm{Vavilov}}$ obtained from 
the fit were used to produce the calibration plot 
presented in Fig.~\ref{calib_nuance}.  
From the hit multiplicity of the hadron shower in any event in ICAL, 
the hadron energy can be estimated using this information. This can then further 
be used to reconstruct the energy and direction 
of the incident neutrinos. The details of this 
reconstruction will be discussed elsewhere \cite{MMDL}. 

\newpage
\section{Summary and Conclusions} 
\label{sec:concl} 
 
The ICAL detector is expected to be a good tracking detector for the muons 
produced in charged-current atmospheric $\nu_{\mu}$ events. However, in 
order to reconstruct the neutrino energy, the estimation of the energy 
of the hadrons produced in such interactions is crucial. Recently, a 
physics simulation study of neutrino mass hierarchy determination with 
ICAL has shown that the inclusion of hadron energy information appreciably 
improves the sensitivity of ICAL to this important issue that is still
unresolved \cite{CG}. Reconstructing the hadron energy and directions
is also the only way to investigate the neutral current events, which
can provide information on tau neutrinos, and on phenomena like
active-sterile oscillations. It is therefore important to characterize 
the behaviour of hadrons in the detector.

A GEANT4-based simulations framework of the ICAL detector has been employed 
to explore the energy calibration of hadrons from the hadron hit pattern 
and to obtain their energy resolutions in the region of interest to 
atmospheric neutrino oscillations. The results of this simulation would be 
crucial for the studies that analyze the reach of the ICAL detector for 
determining the neutrino mixing parameters.
 
The hadron events of interest in the ICAL detector primarily contain
charged pions. The hit pattern of pions and kaons in the detector is 
similar; hence it is not possible to separate different hadrons in the 
detector. Similarly, neutrino-nucleus interactions produce events with 
multiple hadrons in the final state (generated by the NUANCE neutrino 
generator), whose energies cannot be reconstructed individually.  However, 
the {\em total energy} deposited in hadrons can be determined by a 
calibration against the hit multiplicity of hadrons in the detector. 

The hit patterns in single and multiple hadron events are roughly similar, 
and may be described faithfully by a Vavilov distribution.
Analyses, first with fixed-energy pions, and later with a mixture of 
hadrons from atmospheric $\nu_{\mu}$ interaction events, show that 
a hadron energy resolution in the range 85\% (at 1 GeV) -- 36\% (at 15 GeV) 
is obtainable for hadrons produced in charged-current neutrino interactions. 
The parameters of the Vavilov fit presented here as a function of hadron energy
can be used for simulating the hadron energy response of the detector,
in order to perform physics analyses that need the hadron 
energy resolution of ICAL. We also present the calibration for the energy of the hadron 
shower as a function of the hit multiplicity. This analysis will be improved upon by
incorporating edge effects and noise in a later study, after data from 
the prototype detector is available. 

The results in this paper will allow us to reconstruct the total visible
energy in NC events. Combined with the information on the muon energy
and direction in the CC events, it will allow one to reconstruct the 
total neutrino energy in the CC events. ICAL will be one of the largest
neutrino detectors sensitive to the final state hadrons in neutrino
interactions, and its potential for extracting hadronic information 
needs to be fully exploited.

\section{Acknowledgements}  
 
This work is a part of the ongoing effort of the INO collaboration 
to study the physics potential of the proposed ICAL detector at INO. 
We would like to thank Gobinda Majumder and Asmita Redij for 
their contribution in developing the simulation framework for ICAL,
Tarak Thakore and Meghna K K for discussions on GEANT4 and the NUANCE 
event generator, and Prafulla Behera, Naba Mondal, as well as all the 
other members of INO collaboration for the many constructive suggestions 
and criticisms. We also thank the Department of Atomic Energy (DAE) and 
the Department of Science and Technology (DST), Government of India, 
for financial support. One of us (DK) would like to thank the Council 
for Scientific and Industrial Research (CSIR), India for financial support.


\appendix 
\numberwithin{equation}{section} 

\section{The Vavilov probability distribution function} 
\label{sec:appA} 
 
As has been observed from Fig.~\ref{histocomp}, the distribution of hit 
multiplicity in the detector, obtained for fixed-energy charged pions 
(and hadrons in general) is asymmetric, particularly at lower energies. 
The Vavilov probability distribution function 
is found to be a suitable one to represent the hit distributions. 
 
The Vavilov probability density function in the standard form is defined by 
\cite{vavilov}
\begin{equation} 
P(x; \kappa, \beta^2) =  
  \frac{1}{2 \pi i}\int_{c-i\infty}^{c+i\infty} \phi(s) e^{x s} ds \; , 
\label{app1} 
\end{equation} 
where  
\begin{equation} 
\phi(s) = e^{C} e^{\psi(s)},~~~C = \kappa (1+\beta^2 \gamma )  \; ,
\label{app2} 
\end{equation} 
and   
\begin{equation} 
\psi(s)= s \ln \kappa + (s+\beta^2 \kappa) 
\cdot \left [ \int \limits_{0}^{1} 
\frac{1 - e^{-st/\kappa}}{t} \,d t~-~ \gamma \right ] 
- \kappa \, e^{-s/\kappa} \; ,
\label{app3} 
\end{equation} 
where $\gamma = 0.577\dots$ is the Euler's constant. 
The parameters mean and variance ($\sigma^2$) of the distribution in 
Eq. (\ref{app1}) are given by  
\begin{equation} 
\text{mean}= \gamma -1 - \ln\kappa-\beta^2;~~~  
\sigma^2 = \frac{2-\beta^2}{2\kappa}.  
\end{equation} 
For $\kappa\le 0.05$, the Vavilov distribution may be approximated by the
Landau distribution, while for $\kappa\ge10$, it may be approximated by
the Gaussian approximation, with the corresponding mean and variance. 
 
We have used the Vavilov distribution function $P(x; \kappa, \beta^2)$ 
defined above, which is also built into ROOT, as the basic distribution 
for the fit. However the hit distribution itself is fitted to the 
modified distribution 
$\left({\rm P}_4/{\rm P}_3 \right)\;
P( ({\rm x}- {\rm P}_2)/{\rm P}_3; 
~{\rm P}_0,~{\rm P}_1)$,  
to account for the x-scaling (${\rm P}_3$), normalization ${\rm P}_4$ and the 
shift of the peak to a non-zero value, ${\rm P}_2$.
Clearly ${\rm P}_0=\kappa$ and ${\rm P}_1 = \beta^2$. 
The modified mean and variance are then 
\begin{equation} 
\mbox{Mean}_{\rm Vavilov} = (\gamma~-~1~-~\ln {\rm P}_0- {\rm P}_1)
~{\rm P}_3+ {\rm P}_2\; , \quad 
\sigma^2_{\rm Vavilov}= \frac{(2- {\rm P}_1)}{2 {\rm P}_0} {\rm P}_3^2 \; . 
\end{equation}  
 

\end{document}